\documentclass[research]{fcs}
\usepackage{layout}
\usepackage{subfig}
\usepackage{hyperref}

\usepackage{graphicx}
\usepackage{enumitem}
\usepackage{amsmath} 
\usepackage{algorithm}
\usepackage{algpseudocode}
\usepackage{amsmath} 
\usepackage{amssymb} 
\usepackage{cite}

\usepackage{xspace}
\newcommand{\system}{AcOrch\xspace}

\newcommand{\notecolor}[1]{\textcolor{black}{#1}}

\title{\system: Accelerating Sampling-based GNN Training under CPU-NPU Heterogeneous Environments}

\author[1]{Kefu~CHEN}
\author[1]{Xin~AI}
\author[2]{Qiange~WANG}

\author[1,+]{Yanfeng~ZHANG}

\author[1]{Ge~YU}
\address[1]{Department of Computer Science and Engineering, Northeastern University, Shenyang 110167, China}
\address[2]{School of Computing, National University of Singapore, Singapore 119077, Singapore}

\corremail{zhangyf@mail.neu.edu.cn}
\fcssetup{
  received = {June 25, 2025},
  accepted = {October 11, 2025},
  issue = {5},
  volume = {21},
  publish-year = {2027},
  publish-month = {May},
}

\begin{abstract}

Graph Neural Networks (GNNs) have achieved remarkable success in various applications. Sampling-based GNN training, which conducts mini-batch training on sampled subgraphs, has become a promising solution for large-scale graphs. Given the resource-intensive nature of sampling-based GNN training, Neural Processing Units (NPUs), such as the Ascend AI processor, offer a promising alternative due to their high throughput and energy efficiency, making them well-suited for GNN workloads. However, the multi-stage nature of sampling-based training, which involves subgraph sampling, feature gathering, and model training, with different resource requirements and computation volume. This requires careful coordination to fully utilize the heterogeneous computation resources of CPUs and NPUs. In this work, we present AcOrch, a sampling-based GNN training system optimized for CPU-NPU heterogeneous platforms. AcOrch offers fine-grained task orchestration and adopts a two-level pipelined execution model to overlap sampling, gathering, and training. It analyzes the heterogeneous compute features of NPUs and maps tasks to AI Cube (AIC) units, AI Vector (AIV) units, and CPU cores accordingly. 
Moreover, the two-level pipeline enables overlapping execution not only between the CPU and NPU, but also among different types of compute units within the NPU (e.g., AIC and AIV units), thereby maximizing the utilization of available resources.
Experiments on an Ascend 910B AI processor show that AcOrch achieves an average speedup of 2.31$\times$ over the state-of-the-art NPU-native graph learning system, MindSporeGL.
\end{abstract}
\keywords{GNN training; Ascend AI processor; task orchestrating}

\begin{document}

\section{Introduction}
%



Graph Neural Networks (GNNs) have become a fundamental technique for learning on graph-structured data, enabling breakthroughs in applications such as recommendation, fraud detection, and molecular property prediction\cite{neuorch,gnnworkssocial,neutronstar,dgl,XGCN,gcn,turboGNN,gnnspmethod_ieeejas_2022,gnnsurvey_tnn21,fcs-gnnsurvey,fcs-dynamicgnnsurvey}. As real-world graphs grow larger and more complex, sampling-based training methods have emerged as the mainstream approach for scaling GNNs\cite{neuorch,turboGNN,sancusgnn_ijcai_2023,gnnspcheat-tcss24,gnnsphetero-mbec24,gnnspfuse-cluster24,gnnspcase_sigops_2021}. By constructing mini-batch subgraphs dynamically, these methods significantly reduce memory consumption and computational cost compared to full-graph training.

\begin{figure}[!t]
  \centering
  \includegraphics[width=1\linewidth]{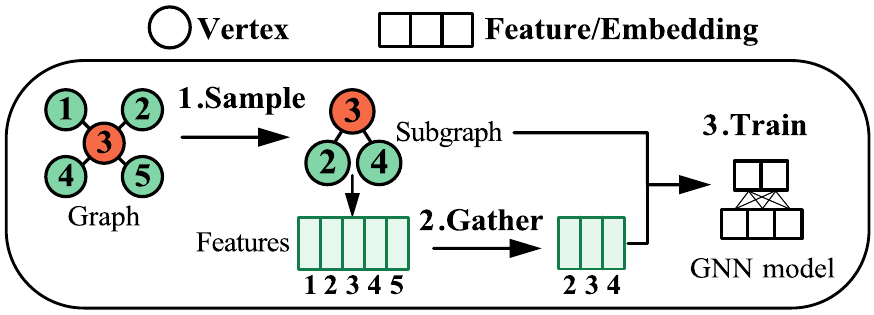}
  \caption{Workflow of sampling-based GNN training.}
  \vspace{-0.1in}
  \label{fig:workflow}
\end{figure}

To conduct mini-batch train on sampled subgraphs, the process typically consists of three steps: subgraph sampling (sampling), feature gathering (gathering), and model training (training)\cite{neuorch,turboGNN,EfficientSampling_train,Zhu_2024,alimohammadi2023localgraphlimitsperspective,10.1145/3492321.3519557,10.1145/3534678.3539038,10.1145/3589311,10.14778/3476249.3476264,chen2018fastgcnfastlearninggraph,cpu-fpga,gnnworkssocial,huang2024freshgnnreducingmemoryaccess,fey2021gnnautoscalescalableexpressivegraph,mohoney2021mariuslearningmassivegraph,XGCN,WholeGraph-sc22,dgl}, as shown in \cref{fig:workflow}. 
In the sampling step, the system selects a set of target nodes and builds their $K$-hop subgraphs based on a given neighbor sampling strategy\cite{sagnn-iscas23,cpu-fpga,gnnspcheat-tcss24,gnnsphetero-mbec24,gnnspfuse-cluster24,gnnsplossspeed-ispa24}. The gathering step retrieves the input features of all vertices involved in the sampled subgraph\cite{gnnfeaturechache-22,turboGNN,chen2018fastgcnfastlearninggraph,cpu-fpga}. Finally, the training step performs a forward and backward pass over the subgraph using the GNN model and updates the corresponding parameters\cite{neuorch,gnnsphetero-mbec24,gnnspfuse-cluster24,neutronstar,XGCN}. The irregular data access patterns and multi-stage processing pipelines of sampling-based methods place high demands on computing resources, necessitating efficient task scheduling and high-throughput execution to fully realize their performance potential.



In recent years, Neural Processing Units (NPUs) have become increasingly popular in modern AI infrastructures, widely adopted in data centers and edge devices for accelerating deep learning workloads\cite{ascend_scalable,Ascend_bench_2}. Representative NPUs include Google’s TPU, Intel’s Habana Gaudi, and Huawei’s Ascend AI processors\cite{TPUv4,Gaudi2,Ascend_bench_4,dhar2024ascendcc,tpu_corr25,ascend_scalable}. These specialized accelerators offer high parallelism, low energy consumption, and strong support for matrix operations, making them a compelling alternative to traditional GPU-based solutions. Given their strengths, NPUs are well-suited for accelerating sampling-based GNN training, which involves both computation-intensive and memory-intensive operations. 
Among them, the Ascend AI processor serves as a representative example, which consists of two types of compute units: AI Cube (AIC) and AI Vector (AIV)\cite{Ascendllm_arxiv_2024}. AIC is responsible for dense
tensor computation and provides the primary computing power of the NPU, while AIV is dedicated to vector-based operations.
This heterogeneous design enables fine-grained mapping of different computational tasks to the most appropriate hardware units, offering opportunities to improve performance and resource efficiency in GNN training.

\begin{figure}[!t]
  \centering
  \includegraphics[width=1\linewidth]{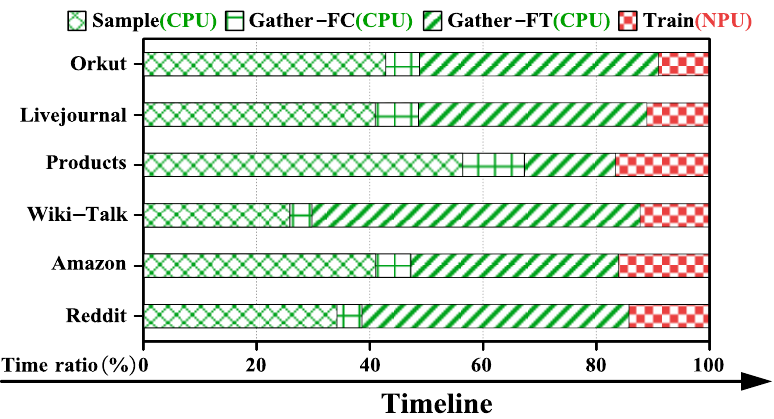}
  \caption{Active timeline of sampling-based GNN training in an iteration across different datasets (proportion of total iteration time consumed by each step). Gather-FC and Gather-FT represent feature collection and feature transfer of the gathering step, respectively.}
  \label{fig:motivation}
\end{figure}

Despite the significant architectural advantages of NPUs, an efficient sampling-based GNN training still requires specialized framework support\cite{dhar2024ascendcc,ascend_scalable,Ascend_bench_1,Ascend_bench_2,Ascend_bench_3,Ascend_bench_4,Ascendllm_arxiv_2024,ascend-cs24}. 
As current native GNN training framework for the Ascend platform, e.g., MindSporeGL\cite{Ascend_bench_1}, generally adopts the traditional task orchestration strategy, assigning sampling and gathering to the CPU while assigning only the model training step to the Ascend AI processor, which significantly limits system performance. 
As shown in \cref{fig:motivation}, with a 2-layer GraphSAGE\cite{reddit} model and a batch size of 4096, experiments using MindSporeGL on six real-world graph datasets (Reddit\cite{reddit}, Amazon\cite{orkut_amazon}, Wiki-Talk\cite{10.1145/1772690.1772756}, Products\cite{products}, Livejournal\cite{Livejournal}, and Orkut\cite{orkut_amazon}) demonstrate that sampling and gathering occupy 83\% to 91\% of the total training time per iteration. During this period, the Ascend AI processor remains largely idle while waiting for data preparation on CPU, resulting in underutilized computational capacity.
In addition, within the Ascend AI Core\cite{dhar2024ascendcc,ascend_scalable,Ascend_bench_1,Ascend_bench_2,Ascend_bench_3,Ascend_bench_4,Ascendllm_arxiv_2024,ascend-cs24}, MindSporeGL also demonstrates inefficiency during the training phase by assigning graph aggregation, a memory-bound operation, to the AIV, and reserving the AIC solely for dense tensor computation. In practice, most training time is consumed by aggregation tasks, with the majority of operations executed on the AIV, leaving the more powerful AIC underutilized throughout the execution. This imbalance in resource allocation prevents the system from fully exploiting the computational potential of the Ascend architecture.


To address the above limitations
of existing frameworks, we propose \system, a sampling-based GNN training system optimized for CPU-NPU heterogeneous environments. \system introduces a fine-grained task orchestration strategy that aligns GNN computation with the architectural characteristics of the Ascend AI processor, enabling balanced resource utilization and efficient execution across various compute units. 
It carefully partitions the workload by assigning sampling to both AIV and CPU, gathering to AIV, and graph aggregation operations and neural network operations, the main operations in training phase, to AIC. This division fully leverages the strengths of each component. A dynamic cost model guides the task allocation, and a multi-producer–single-consumer queue design ensures high-throughput parallel sampling execution. In addition, \system employs a two-level pipelined execution model to maximize parallelism. The first-level pipeline overlaps CPU-based sampling with NPU-based gathering and training, effectively masking communication latency. The second-level pipeline overlaps AIV-based memory operations and AIC-based computation within the NPU using asynchronous queues and double buffering. This two-level pipeline design ensures continuous task execution across devices and within the NPU, substantially improving overall system throughput and resource utilization.

The main contributions of this paper are as follows:
\begin{itemize}

    \item We analyze the resource utilization bottlenecks of sampling-based GNN training on CPU-NPU platforms, identifying key inefficiencies in existing task scheduling strategies.

    \item We propose \system, a sampling-based GNN training system with workload-aware task orchestration and a two-level pipelined execution model that improves hardware utilization across CPU, AIC, and AIV.
    

    \item We implement \system on the Ascend 910B processor and conduct extensive experiments. Our analysis shows that it achieves an average 2.31$\times$ speedup and a 52.63\% improvement in AIC utilization over the state-of-the-art NPU-native baseline, MindSporeGL.

\end{itemize}

\section{Background}

\subsection{Ascend AI processor}

\begin{figure}[!t]
  \centering
  \includegraphics[width=1\linewidth]{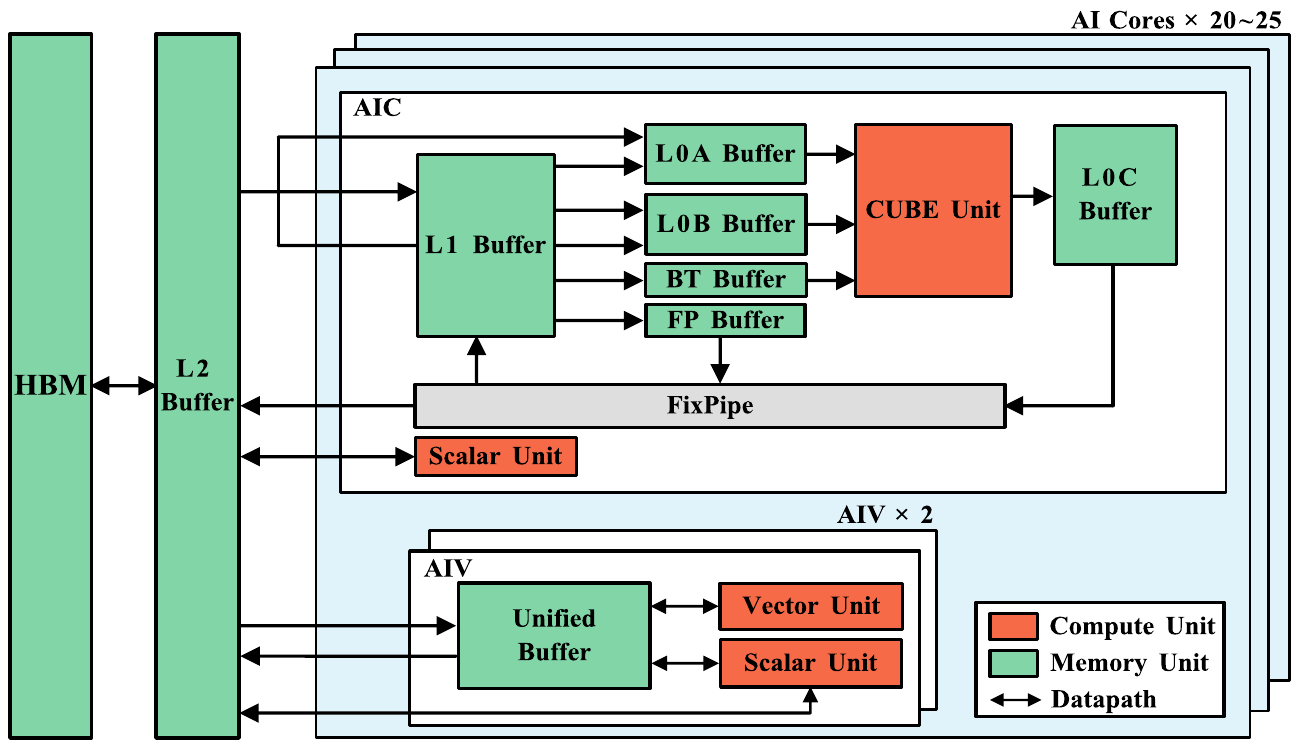}
  \caption{Decoupled architecture of Ascend AI processor.}
  \label{fig:Ascend}
\end{figure}

Huawei Ascend AI processors are based on the self-developed DaVinci architecture and are specifically designed for high-performance AI computing scenarios\cite{ascend_scalable,Ascend_bench_1,Ascend_bench_2,Ascend_bench_3,Ascend_bench_4,ascend-martrix,Ascendllm_arxiv_2024}. This architecture adopts a decoupled design, dividing the AI Core into two independent components: AI Cube (AIC) and AI Vector (AIV)\cite{Ascendllm_arxiv_2024}, which are responsible for high-throughput matrix multiplication and vector operations, respectively. The AIC functions similarly to Nvidia's Tensor Core\cite{Ascendllm_arxiv_2024,ascend_scalable,Ascend_bench_1,GPGPU-iscas24}, and provides the primary computing power of the NPU. Taking the Ascend 910B as an example (see \cref{fig:Ascend}), each AI Core contains one AIC and two AIVs. Both the AIC and AIV are equipped with independent scalar unit, enabling them to load and execute their own code, thereby achieving the decoupling of matrix and vector computations. Data transfer between the AIC and AIV is facilitated by the L2 buffer, with data movement managed by the Memory Transfer Engine (MTE)\cite{Ascend_bench_2,Ascend_bench_3,Ascendllm_arxiv_2024}. 
Compared to GPUs, Ascend AI processors have larger core granularity, making load balancing between cores even more critical\cite{Ascendllm_arxiv_2024}.
Multi-level buffers, global memory, and on-chip high-speed interconnects collectively ensure efficient data exchange. The system software orchestrates the coordinated operation of all computing units, achieving physical isolation and logical collaboration. 
This effectively enhances hybrid computing efficiency and hardware resource utilization\cite{Ascend_bench_1,ascend_scalable}.

\subsection{Graph neural networks}

Graph Neural Networks (GNNs) are deep learning models designed for graph-structured data. Each layer of a GNN consists of an aggregation phase and an update phase, where nodes aggregate information from their neighbors and update their own embeddings accordingly\cite{gnnsurvey_tnn21,neutronstar,neuorch,turboGNN,gnnspcase_sigops_2021,gnnfeaturechache-22,gnnhbmsample-hpec24,gnnspcheat-tcss24,gnnspfuse-cluster24,gnnsphetero-mbec24,gnnsplossspeed-ispa24,gnnspmethod_ieeejas_2022,hongtu-corr23}. For a GNN with $L$ layers, at the $l$-th layer, each node $v$ aggregates the embeddings of itself and its in-neighbors from the previous layer, which can be formulated as:
\begin{equation}
a_v^{(l)} = \text{AGGREGATE}^{(l)} \left( \left\{ h_u^{(l-1)} \mid \forall u \in N_{in}(v) \cup \{v\} \right\} \right),
\end{equation}
where $N_{in}(v)$ denotes the set of in-neighbors of node $v$, and $h_u^{(l-1)}$ represents the embedding of node $u$ at the $(l-1)$-th layer. Through the aggregation operation (such as sum, mean, max, or LSTM), node $v$ can effectively integrate the structural and feature information from its neighborhood\cite{neuorch,reddit,gnnsurvey_tnn21,turboGNN}.
Subsequently, node $v$ feeds the aggregated result $a_v^{(l)}$ into the update function, which, combined with a learnable weight matrix $W^{(l)}$, yields the new embedding at the current layer:
\begin{equation}
h_v^{(l)} = \text{UPDATE}^{(l)} \left( W_v^{(l)}, a_{v}^{(l)} \right).
\end{equation}


By stacking multiple layers, each node is able to capture multi-hop structural and feature information, transforming its feature vector into a low-dimensional embedding. These embeddings can then be utilized for downstream tasks such as node classification and link prediction\cite{turboGNN,gnnhbmsample-hpec24,gnnspcheat-tcss24,gnnsphetero-mbec24,cpu-fpga}.

\begin{figure}[!t]
  \centering
  \includegraphics[width=1\linewidth]{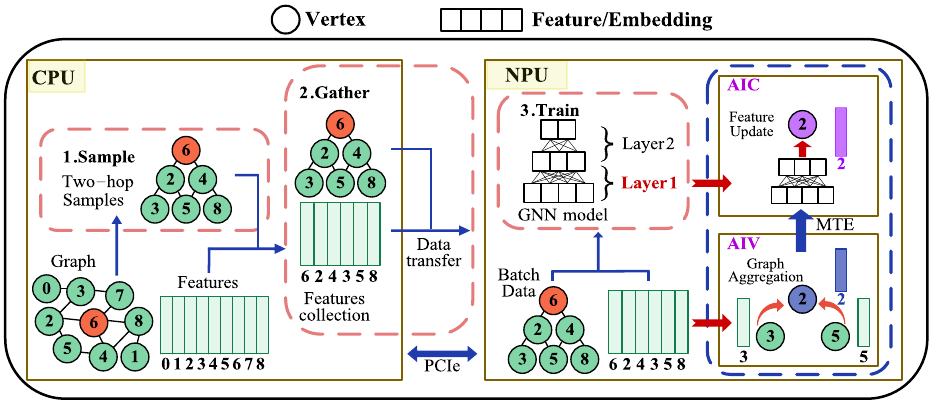}
  \caption{An example of sampling-based training for a two-layer GNN on MindsporeGL in a CPU-NPU heterogeneous environment, where vertex 6 is a training vertex with a ground-truth label.}
  \label{fig:3fanshi}
\end{figure}

\subsection{Sampling-based mini-batch GNN training on CPU-NPU}


Sampling-based mini-batch GNN training divides training nodes into batches and dynamically samples local subgraphs centered on target nodes, effectively reducing the computational and storage burden per iteration\cite{neuorch,neutronstar,10.1145/1772690.1772756,turboGNN,cpu-fpga,gnnspcheat-tcss24,gnnsphetero-mbec24,gnnspfuse-cluster24,gnnsurvey_tnn21,Quiver_corr_2023}. Taking the native Ascend GNN training framework MindSporeGL as an example (see \cref{fig:3fanshi}), consider node 6 as a training node with a ground-truth label. 
The training process consists of three steps: subgraph sampling (sampling), feature gathering (gathering), and model training (training). In the \textbf{sampling} stage, the CPU samples the neighbors of target node 6 according to the sampling strategy and constructs a multi-hop subgraph. In the \textbf{gathering} stage, the CPU gathers the input features of node 6 and its sampled neighbors from CPU memory and transfers the subgraph and feature data to the NPU via PCIe. In the \textbf{training} stage, the NPU executes the forward and backward propagation of the GNN model and updates the parameters\cite{neuorch,turboGNN,mohoney2021mariuslearningmassivegraph,10.1145/3589311,gnnspcheat-tcss24,gnnspfuse-cluster24,gnnsphetero-mbec24,gnnsplossspeed-ispa24}.
During GNN training execution, graph aggregation and neural network (NN) operations within the NPU are handled by the AIV and AIC, respectively.
Taking node 2 at layer 1 as an example, the initial feature vectors of its neighbors, node 3 and node 5, are first aggregated in the AIV. The aggregated result is then processed in the AIC through NN operations to generate a new embedding for node 2.


\subsection{Existing CPU-GPU GNN training frameworks}




Existing works has proposed a variety of task orchestration strategies for sampling-based GNN training in CPU-GPU heterogeneous environments, typically following the \textbf{sample-gather-train} paradigm with different combinations of task assignments across CPU and GPU resources \cite{neuorch,turboGNN,EfficientSampling_train,Zhu_2024,alimohammadi2023localgraphlimitsperspective,10.1145/3492321.3519557,10.1145/3534678.3539038,10.14778/3476249.3476264,cpu-fpga,gnnworkssocial,fey2021gnnautoscalescalableexpressivegraph,mohoney2021mariuslearningmassivegraph,huang2024freshgnnreducingmemoryaccess,gnnspcheat-tcss24,gnnsphetero-mbec24,gnnsplossspeed-ispa24,gnnspfuse-cluster24,chen2018fastgcnfastlearninggraph}. 
However, due to the physical coupling of heterogeneous resources, either the CPU or GPU inevitably needs to handle at least two steps on a single machine, resulting in either inefficient CPU processing or GPU resource contention. The former arises from the CPU’s irregular memory access patterns and PCIe bandwidth limitations, creating a bottleneck in data preparation; the latter stems from contention for computing resources or memory within the GPU, leading to reduced utilization\cite{10.1145/3589311,10.1145/3534678.3539038,neuorch,turboGNN,cpu-fpga}. Traditional step-based orchestration struggles to balance cross-device workloads and cannot fully exploit the heterogeneous hardware potential. We identify imbalanced resource allocation as the main performance bottleneck and emphasize the need for finer-grained cross-device scheduling to improve efficiency.

\subsection{Opportunities and challenges of GNN training on NPUs}

The diverse computation patterns of sampling-based GNN training motivate the need for more flexible and specialized hardware support.
NPUs introduce new opportunities for accelerating sampling-based GNN training due to their inherently heterogeneous and physically decoupled multi-core architecture\cite{ascend_scalable,ascend-martrix,Ascend_bench_1,Ascend_bench_2,Ascend_bench_3}. Unlike traditional CPU-GPU systems, NPUs offer multiple specialized compute units that can better match the distinct computational characteristics of different stages. 
Taking the Ascend AI processor as an example, it provides fine-grained resource partitioning through decoupled components like the AIV and AIC units. 
This separation enables independent scheduling and parallel execution within the NPU, reducing resource contention and supporting flexible pipeline scheduling. These capabilities are particularly advantageous for overlapping multi-stage GNN training workflows, making Ascend an ideal platform for exploring efficient orchestration strategies in heterogeneous GNN training.


Nevertheless, such architectural diversity also introduces unique system-level challenges. The heterogeneous compute units within the Ascend AI processor necessitate highly adaptive and task-specific scheduling, as opposed to the relatively uniform resource management in GPU-based systems. Existing CPU-GPU orchestration strategies, which assume that each device contains homogeneous compute resources, are inadequate for addressing the intricate heterogeneity and decoupled execution model of NPUs. As a result, directly transferring these strategies to NPU platforms often results in inefficient resource utilization and new performance bottlenecks. To fully exploit the advantages of NPUs, it is thus essential to develop customized task partitioning and scheduling methods that align with their distinct architectural characteristics.


\section{Analysis of task orchestrating methods on CPU-NPU}


\begin{figure*}[!t]
  \centering
  \includegraphics[width=1\linewidth]{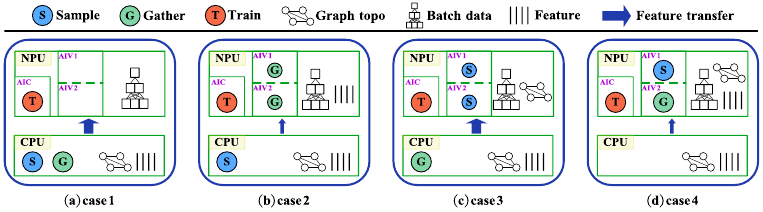}
  \caption{Illustration of four typical task orchestration methods. 
  (a) Case 1: Assigning both sampling and gathering to the CPU. 
  (b) Case 2: Assigning sampling to the CPU and gathering to the AIV. 
  (c) Case 3: Assigning sampling to the AIV and gathering to the CPU.
  (d) Case 4: Assigning both sampling and gathering to the AIV.
  The width of the arrow indicates data transfer volume.}
  \vspace{-0.1in}
  \label{fig:step_orch}
\end{figure*}

\begin{figure}[!t]
  \centering
  \includegraphics[width=1\linewidth]{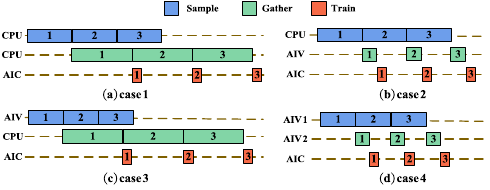}
  \caption{Examples of multi-stream pipeline design under the four typical task orchestration method. The width of each block is proportional to its execution time.}
  \vspace{-0.15in}
  \label{fig:pipeline}
\end{figure}


On heterogeneous platforms, sampling-based GNN training typically adopts a step-wise decoupling approach, with each step assigned to a device. The mapping of sampling, gathering, and training across heterogeneous hardware directly determines overall system performance\cite{WholeGraph-sc22,neuorch,turboGNN}. 
Compared with GPUs, the Ascend AI processor further subdivides computational resources at the physical level, primarily into AIC for neural network computation and AIV for flexible vector processing, providing greater flexibility for task scheduling\cite{ascend_scalable,Ascend_bench_1,Ascend_bench_2,Ascend_bench_3}. Since the AIC is optimized for compute-intensive neural network operations but not suited for memory-bound tasks such as sampling and feature gathering, the training step of existing systems is generally assigned exclusively to the AIC.
Consequently, the scheduling of sampling and gathering tasks between the CPU and AIV yields four possible combinations, as illustrated in \cref{fig:step_orch}. Next, we introduce each case in detail and analyze their performance through experiments.




\begin{enumerate}[leftmargin=0pt,itemindent=2em,label=\arabic*)]
\item Case 1: Assigning both sampling and gathering to the CPU.
As shown in \cref{fig:step_orch} (a), during each iteration, the CPU first performs neighbor sampling, then gathers the required node features, and finally transfers the prepared graph topology and feature data in batches to the NPU for training. Throughout the data preparation phase, the NPU remains idle and only begins computation after receiving the complete batch of data from the CPU. 
Since all data preparation tasks are handled by the CPU, the overall system performance is constrained by the CPU’s throughput and PCIe bandwidth, resulting in significant pipeline bubbles and extremely low hardware resource utilization, as illustrated in \cref{fig:pipeline} (a).


\item Case 2: Assigning sampling to the CPU and gathering to the AIV.
As shown in \cref{fig:step_orch} (b), the CPU first performs neighbor sampling and then sends the sampled subgraph topology to the NPU. The AIV is responsible for gathering the corresponding node features from the NPU’s memory, and together with the subgraph topology, delivers them to the AIC for training.
By offloading feature collection to the AIV and storing all features or caching high-frequency features in the NPU’s global memory, the PCIe communication bottleneck during feature transfer is significantly reduced. Although this approach achieves a balanced mapping between tasks and hardware resources, the random memory access pressure on the CPU during sampling remains substantial, becoming the main bottleneck. Prolonged data preparation delays subsequent training, and the parallel capabilities of AIV and AIC are not fully utilized, as illustrated in \cref{fig:pipeline} (b).


\item Case 3: Assigning sampling to the AIV and gathering to the CPU. 
As shown in \cref{fig:step_orch} (c), the AIV first completes graph sampling, after which the CPU collects the corresponding node features from main memory and transfers the processed feature data in batches to the NPU for training by the AIC.
In this mode, the gathering step is again constrained by the PCIe bandwidth, becoming the dominant system bottleneck and limiting AIC utilization. Furthermore, although the AIV supports SIMD instructions to enhance parallel processing, its flexibility is inferior to the CPU. When faced with a large volume of irregular random accesses in sampling, effective acceleration is difficult to achieve, and the sampling bottleneck persists, cannot efficiently utilize the pipeline, as illustrated in \cref{fig:pipeline} (c).


\item Case 4: Assigning both sampling and gathering to the AIV.
As shown in \cref{fig:step_orch} (d), the graph topology and feature data are cached in the NPU, and the AIV1 first performs subgraph topology sampling, followed by the AIV2 collecting the corresponding node features from NPU memory. Finally, both are sent to the AIC for training.
This approach better leverages the NPU’s internal parallel hardware capabilities. However, the AIV must simultaneously handle intensive random access and feature collection, resulting in significant hardware pressure. Although the Ascend decoupled architecture equips each AI Core with an AIC and multiple AIVs, partially alleviating AIV resource contention, sampling on the AIV remains a performance bottleneck. AIC utilization remains low, and CPU resources are left idle for extended periods, leading to imbalanced overall hardware resource scheduling, as illustrated in \cref{fig:pipeline} (d).
\end{enumerate}



\begin{figure}[!t]
  \centering
  \includegraphics[width=1\linewidth]{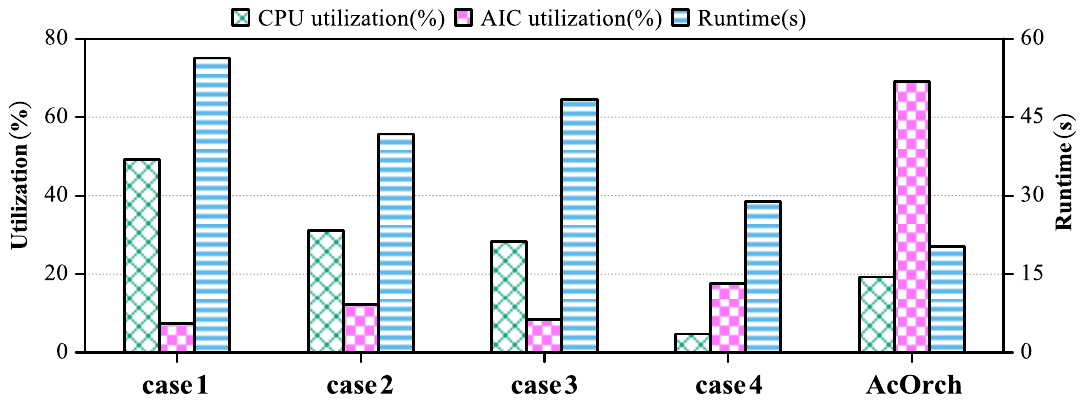}
  \caption{Comparison of resource utilization and per-epoch execution time for the existing task orchestrating methods.}
  \label{fig:tradition_orch}
\end{figure}

\cref{fig:tradition_orch} illustrates the impact of different task scheduling strategies on resource utilization and training performance of a CPU-NPU heterogeneous platform using the Reddit dataset and a two-layer GraphSAGE model. As the AIC is the core computing unit of the Ascend AI processor, we primarily focus on AIC utilization. Experimental results show that concentrating sampling and gathering on the CPU (MindSporeGL) results in the longest epoch time and the lowest AIC utilization (only 7.41\%), with training bottlenecked by data preparation. As tasks are gradually migrated to the NPU, the CPU bottleneck is alleviated and training efficiency improves, but AIC utilization remains low, with an average of only 11.44\% across the four strategies. Among the four strategies, Case 4 achieves the best epoch time and AIC utilization, yet is still constrained by data preparation delays in the AIV sampling step. Consequently, frequent data waits in the pipeline lead to significant AIC idleness. Allocating all tasks to the NPU does not markedly improve AIC utilization, which only reaches 17.74\%, and CPU resources remain underutilized.


The above analysis indicates that step-based coarse-grained task scheduling, regardless of how sampling, gathering, and training are allocated, fails to achieve balanced and efficient resource utilization on CPU-NPU heterogeneous platforms. The sampling step, whether assigned to the CPU or AIV, consistently constitutes a performance bottleneck in the training process, preventing the effective exploitation of AIV and AIC hardware capabilities, resulting in prolonged idleness of AIC computing power and limited system throughput. Moreover, such coarse-grained orchestration lacks autonomous workload balancing. Therefore, there is an urgent need for a fine-grained task orchestration and pipelined scheduling mechanism that can accurately sense the workload of each stage and dynamically optimize cross-device resource allocation, thereby overcoming the sampling bottleneck, fully unleashing the hardware potential of both CPU and NPU, and improving overall system efficiency.

\section{\system}

\subsection{Sample-driven task orchestration}




Although step-based task orchestration can leverage heterogeneous hardware resources, it remains constrained by coarse-grained workload allocation. Once entire stages such as sampling, feature collection, or training are assigned to specific hardware resources, the corresponding workload becomes statically fixed, making dynamic adjustment and fine-grained workload balancing difficult. The analysis in Section 3 shows that regardless of whether the sampling task is assigned to the CPU or AIV, the time spent on collection and training on AIV is significantly less than that of sampling, so the overall system bottleneck always lies in the sampling stage. Therefore, overcoming the bottleneck imposed by sampling is key to improving system throughput. Achieving this requires fine-grained decomposition of the sampling stage and dynamic adjustment capabilities for the decomposed sampling tasks, enabling flexible resource scheduling and efficient workload balancing.



\paragraph{Design insight} The feasibility of fine-grained decomposition of sampling tasks stems from the fact that Ascend AI processors provide physically isolated heterogeneous computing resources such as AIC and AIV. For example, each AI Core in the Ascend 910B integrates one AIC and two AIVs. Compared to homogeneous architectures where allocating multiple stages to the same hardware easily leads to resource contention, the Ascend architecture effectively alleviates this issue. Therefore, in this study, the training node set is further partitioned within the sampling stage, with sampling executed in parallel by CPU and AIV. In conjunction with a multi-producer single-consumer queue and workload-aware task partitioning mechanism, this design not only fully utilizes the computational potential of various resources, but also achieves fine-grained workload balancing, effectively mitigating the impact of the sampling stage on system performance and improving NPU utilization. 


\paragraph{Orchestration methodology}


As a sample-driven task orchestration framework, \system treats sampling as the entry point for coordinating heterogeneous resources, enabling dynamic and efficient execution across the entire training pipeline. Specifically, \system introduces a computation-aware workload partitioning model that dynamically assigns the sampling vertices of each mini-batch to both the CPU and AIV for parallel execution. After sampling, gathering, and training are jointly performed by the AIV and AIC, ensuring efficient collaboration across heterogeneous compute units. On the Ascend 910B processor, the decoupled architecture 
allows sampling and feature gathering to proceed in parallel on different AIVs. This significantly reduces hardware-level resource contention and ensures that the task orchestration aligns well with the underlying architecture.

\begin{figure}[!t]
  \centering
  \includegraphics[width=1\linewidth]{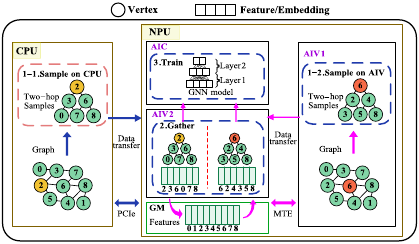}
  \caption{A sampling-based two-layer GNN training example on \system, where vertex 2 and vertex 6 are training vertices after sampling-based vertex partitioning.}
  \label{fig:AcOrch}
\end{figure}


As shown in \cref{fig:AcOrch}, \system first assigns vertices 2 and 6, which have been partitioned from the mini-batch by a workload-aware model, to the CPU and AIV1, respectively, thereby enabling dual-path parallel sampling.
Upon completion, each sampled subgraph is transmitted via PCIe and MTE (Memory Transfer Engine) to AIV2, which independently batches and gathers features from global memory for each path. The sampled subgraphs and their features are then sent to the AIC, where unified scheduling ensures that as soon as a subgraph’s features are ready, forward and backward propagation are performed to complete training. Throughout this process, sampling, gathering, and training proceed independently along two paths; whichever subgraph and features are ready first will be processed first. This parallel and dynamically scheduled task orchestration significantly enhances system resource utilization and overall performance.

\begin{figure*}[t]
  \centering
  \includegraphics[width=1\linewidth]{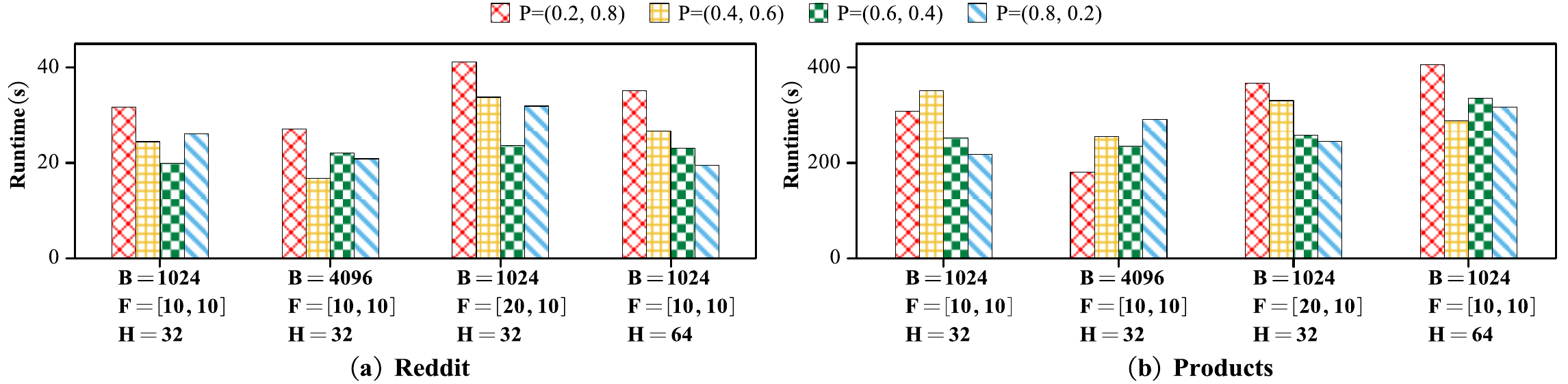}
  \caption{The impact of different AIV/CPU sampling node partition ratios P=(p, 1-p) on overall system execution time (p denotes the proportion assigned to AIV, 1-p denotes the proportion assigned to CPU). B, F, and H are model parameters representing batch size, fanout (number of sampled neighbors), and hidden dimension, respectively.}
  \label{fig:sample_part}
\end{figure*}

\begin{algorithm}[t] 
\caption{Computation-aware workload partition}
\begin{algorithmic}[1]
\label{alg:sample}
\Require Node set $V_{batch}$, workload scores $w(v)$, capability ratio $r = \frac{S_{AIV}}{S_{CPU}}$, performance threshold $T$, previous batch time $t_{prev}$, current batch time $t_{curr}$
\Ensure Assignment result $V_{AIV}$, $V_{CPU}$
\If{$t_{curr} - t_{prev} > T$}
    \State Compute $W \gets \sum_{v \in V_{batch}} w(v)$
    \State Compute $p \gets \frac{r}{1 + r}$
    \State $W_{AIV}^{\text{target}} \gets p \cdot W$, $W_{CPU}^{\text{target}} \gets (1-p) \cdot W$
    \State Sort $V_{batch}$ by $w(v)$ in descending order
    \State Initialize $V_{AIV} \gets \emptyset$, $V_{CPU} \gets \emptyset$
    \State $S_{AIV} \gets 0$, $S_{CPU} \gets 0$
    \For{node $v$ in sorted $V_{batch}$}
        \If{$S_{AIV} < W_{AIV}^{\text{target}}$}
            \State $V_{AIV} \gets V_{AIV} \cup \{v\}$
            \State $S_{AIV} \gets S_{AIV} + w(v)$
        \Else
            \State $V_{CPU} \gets V_{CPU} \cup \{v\}$
            \State $S_{CPU} \gets S_{CPU} + w(v)$
        \EndIf
    \EndFor
    \State Cache $V_{AIV}$, $V_{CPU}$ as the current partition
\Else
    \State Reuse the previous partition $V_{AIV}$, $V_{CPU}$
\EndIf
\Return $V_{AIV}, V_{CPU}$
\end{algorithmic}
\end{algorithm}

\subsection{Computation-aware workload partition}

\paragraph{Sampling workload partition} 


To systematically analyze the impact of the sampling node partition ratio between AIV and CPU on overall system performance, we conducted sensitivity experiments using a two-layer GraphSAGE model with Reddit and Products datasets. The experiments evaluated four partition ratios, p=(0.2, 0.8), (0.4, 0.6), (0.6, 0.4), and (0.8, 0.2), under various model parameter settings including batch size, fanout, and hidden dimension, measuring the overall execution time of the system. For clarity, we denote the four model parameter settings as config 1 to 4, which correspond to (B=1024, F=[10,10], H=32), (B=4096, F=[10,10], H=32), (B=1024, F=[20,10], H=32), and (B=1024, F=[10,10], H=64), respectively.


As shown in \cref{fig:sample_part}, different sampling node partition ratios have a significant impact on system performance. Under the four model parameter configurations, the average speedup between the best and worst system performance resulting from the four sampling node partition ratios is 1.70$\times$ for Reddit and 1.46$\times$ for Products. The optimal partition ratio also changes as model parameters such as batch size vary. For example, in the Reddit dataset, the optimal partition ratio for different configurations is P=(0.6,0.4), P=(0.4,0.6), P=(0.6,0.4), and P=(0.8,0.2), respectively. In addition, under the same model parameter configuration, the optimal partition ratio differs across datasets. For instance, in config 1, the optimal ratio for Reddit is P=(0.6,0.4), while for Products it is P=(0.8,0.2). These results indicate that system performance is jointly influenced by model parameters and dataset characteristics, and that the optimal sampling node partition ratio is highly dynamic and uncertain. Therefore, empirical or static partitioning approaches are insufficient to meet the complex and variable requirements of real-world scenarios. It is essential to develop computation-aware sampling partition strategies that can adaptively adjust according to both data and model characteristics.

To maximize resource utilization and system throughput during heterogeneous sampling, \system adopts a computation-aware workload partitioning strategy. The objective is to achieve workload balancing between heterogeneous compute units, such as CPU and AIV, by distributing the sampling workload of each mini-batch according to the actual computational demand of each vertex and the capabilities of the hardware.

\paragraph{Cost model}

\notecolor{To provide a unified metric for evaluating the computational workload of each node during sampling, we introduce a cost model that considers two components: the node degree $\mathrm{deg}(v)$ and the historical average sampling time $\hat{t}(v)$. A larger $\mathrm{deg}(v)$ indicates that the node needs to access more neighbors, increasing the computational load. $\hat{t}(v)$ records the actual sampling time, and even nodes with the same degree may have different runtime overheads. For any node $v$ in the mini-batch, its workload score is defined as follows: }
 \begin{equation} 
 w(v) = \alpha \cdot \mathrm{deg}_{\mathrm{norm}}(v) + \beta \cdot \hat{t}_{\mathrm{norm}}(v), \end{equation} 
where $\mathrm{deg}_{\mathrm{norm}}(v)$ and $\hat{t}_{\mathrm{norm}}(v)$ denote the standardized values of the degree and historical average sampling time of node $v$, respectively. During the preprocessing stage, multiple random samplings are performed for each selected training node, and their sampling times are recorded.
\notecolor{$\mathrm{deg}_{\mathrm{norm}}(v)$ and $\hat{t}_{\mathrm{norm}}(v)$ exhibit some correlation. We perform principal component analysis (PCA)\cite{pca} to compute the loadings of the first principal component, which reflect the contribution of each variable to the overall variation in node overhead. The absolute values of the loadings are normalized to obtain the weights $\alpha$ and $\beta$.
We normalize the absolute values of these loadings to obtain the weights $\alpha$ and $\beta$. The final workload score $w(v)$ is calculated by weighting $\mathrm{deg}_{\mathrm{norm}}(v)$ and $\hat{t}_{\mathrm{norm}}(v)$ with these coefficients, and we use this score to quantify the sampling overhead of each node.}



Given the node set $V_{batch}$ for a mini-batch, the total workload is computed as $W = \sum_{v\in V_{batch}} w(v)$. 
The hardware processing speeds of the AIV and CPU are denoted by $S_{AIV}$ and $S_{CPU}$, respectively, with the capability ratio defined as $r = S_{AIV} / S_{CPU}$. The values of $S_{AIV}$ and $S_{CPU}$ are obtained in the preprocessing phase by calculating the ratio of the total workload score to the actual processing time for their respective sampling tasks.
The partitioning goal is to allocate nodes to AIV and CPU so that the expected processing time for each device is balanced, that is,
\begin{equation}
\arg\min_{\substack{V_{\text{AIV}} \cup V_{\text{CPU}} = V_{\text{batch}} \\ V_{\text{AIV}} \cap V_{\text{CPU}} = \emptyset}} 
\left( 
\frac{\sum_{v \in V_{\text{AIV}}} w(v)}{S_{\text{AIV}}} - 
\frac{\sum_{v \in V_{\text{CPU}}} w(v)}{S_{\text{CPU}}} 
\right)^2
\end{equation}
this leads to a target workload share of $p = \frac{r}{1 + r}$ for AIV and $1 - p$ for CPU, resulting in
\begin{equation}
W_{AIV}^{\text{target}} = p \cdot W, \quad W_{CPU}^{\text{target}} = (1-p) \cdot W.
\end{equation}

\paragraph{Algorithm description} 

Based on the above cost model, \system\ adopts a greedy partitioning algorithm as illustrated in \cref{alg:sample}. The algorithm first quantifies the workload scores $w(v)$ for all nodes within the mini-batch and calculates the total workload $W$. Subsequently, the capability ratio $r = \frac{S_{AIV}}{S_{CPU}}$, obtained from empirical measurements during the preprocessing stage, is used to dynamically determine the target workload shares for AIV and CPU, which are given by $W_{AIV}^{\text{target}} = \frac{r}{1+r} \cdot W$ and $W_{CPU}^{\text{target}} = \frac{1}{1+r} \cdot W$, respectively.

All nodes are sorted in descending order according to their workload scores $w(v)$, such that nodes with higher computational demand are prioritized in the assignment process. During allocation, the algorithm greedily traverses the sorted nodes; if the current accumulated workload assigned to AIV has not reached its target share, the node is assigned to AIV. Otherwise, the node is assigned to CPU. This process continues until all nodes have been allocated. The partitioning result is cached in the HBM and reused in subsequent mini-batches, while repartitioning is triggered only when the performance metric reaches the preset threshold $T$.


\begin{figure*}[t]
  \centering
  \includegraphics[width=1\linewidth]{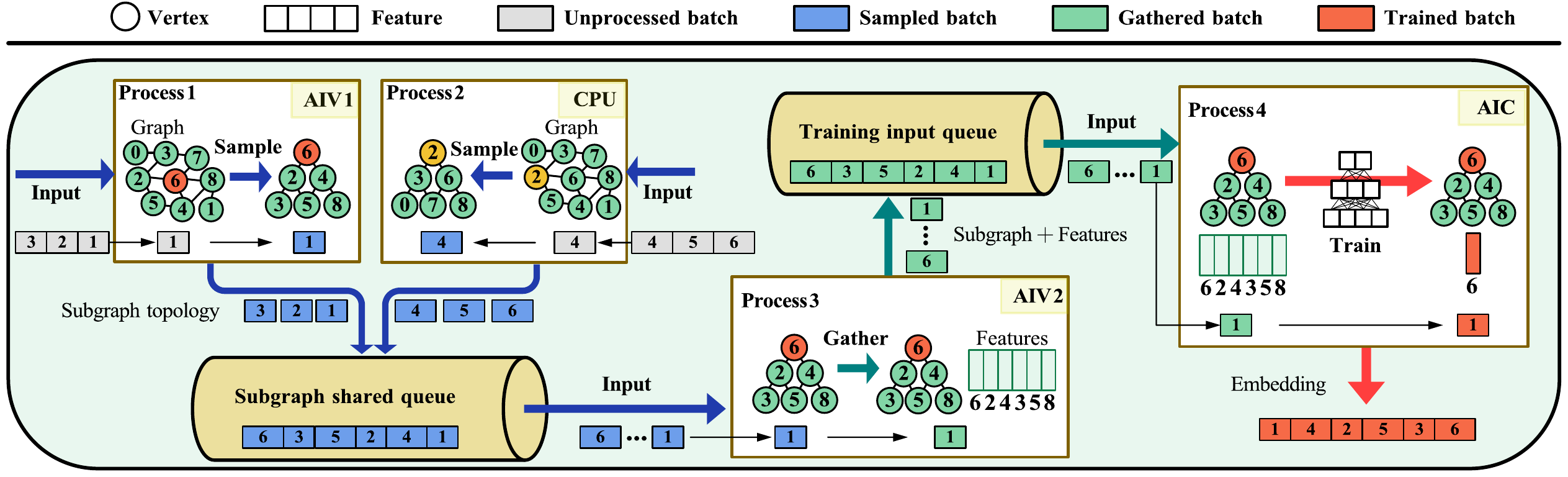}
  \caption{Pipeline architecture with dual-path sampling and shared queue communication for sampling-based GNN training. Each process shows a specific operation, with the corresponding data batch represented by the block below. As the batch progresses through different stages, it changes, and its current processing state is indicated by different colors.}
  \label{fig:shared_que}
\end{figure*}

\subsection{Shared queue communication mechanism}


\cref{fig:shared_que} illustrates the processing flow of a set of mini-batches within the \system pipeline architecture. \system decouples each processing step, executing them in separate processes (process 1 to process 4), and utilizes queues to facilitate inter-process communication. Taking batch 1 (node 6) and batch 4 (node 2) as examples, both initially appear as gray blocks, representing unprocessed target nodes. Sampling is performed in process 1 (AIV1) for batch 1 and in process 2 (CPU) for batch 4. After generating subgraph topologies centered on nodes 6 and 2, the blocks turn blue. Once sampling is completed, all batches are sequentially placed into a shared queue according to the order in which they finish, and then passed as input to process 3 (AIV2). During the gathering stage, batch 1 arrives first, and features of all nodes in its subgraph are collected, causing the block to change from blue to green. At this point, batch 1 contains both the subgraph topology and the node features, while the remaining batches undergo feature collection in the same order. Subsequently, the batches are sent to the training input queue and sequentially forwarded to process 4 (AIC) for training. In the training stage, batch 1 is processed first, and its green block turns red to indicate output, corresponding to the embedding of target node 6; the other batches are processed and output in order as well. Throughout the entire process, each batch transitions through gray, blue, green, and red states at different processing stages, ultimately completing the handling of all batches.

\begin{figure}[t]
  \centering
  \includegraphics[width=0.51\linewidth]{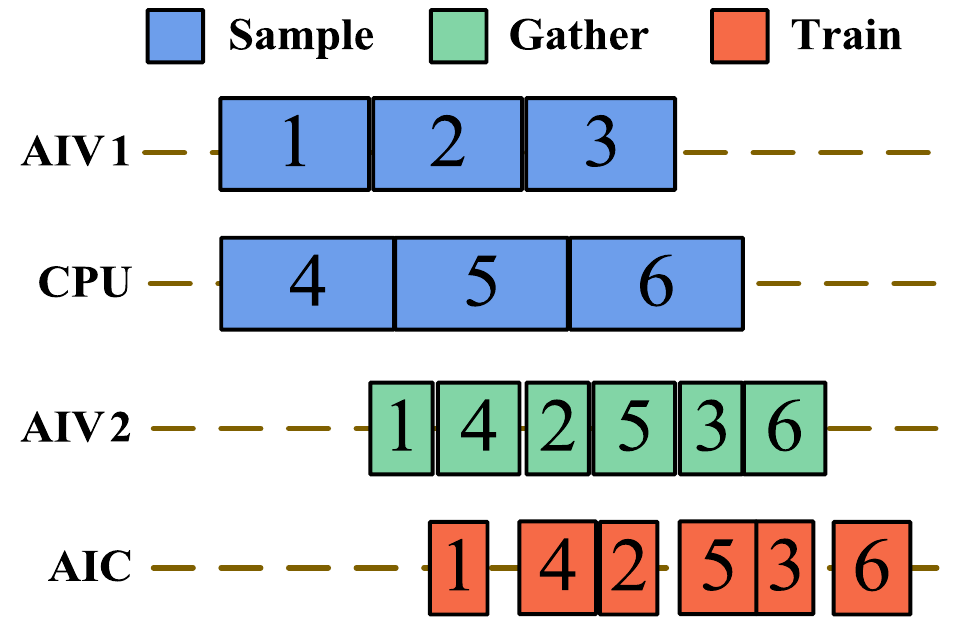}
  \caption{The two-stage pipeline design adopted by \system, the width of the block is proportional to the duration of the phase.}
  \label{fig:AcOrch_pipeline}
    \vspace{-0.1in}
\end{figure}


\system introduces dual-path sampling combined with a shared queue communication mechanism, enabling efficient decoupling and asynchronous parallelism across system stages. For each mini-batch, sampling tasks are partitioned using a computation-aware workload partitioning strategy and executed independently by the CPU and AIV. Sampling processes act as multiple producers, pushing completed subgraph samples to a thread-safe shared queue in real time. The gathering and training processes, as a single consumer, asynchronously retrieve data from the queue, consuming whichever sample is ready first without waiting for all sampling paths. Because gathering and training steps are much faster than sampling step, the queue mechanism smooths out fluctuations between dual sampling paths, preventing pipeline starvation or resource idling and improving throughput and stability. In addition, the decoupling of sampling and training steps through the shared queue allows each stage to progress at its own pace and adapt to runtime variability, further enhancing system robustness and scalability.

\subsection{Two-level pipeline optimization}


\system introduces a two-stage pipeline optimization model, as illustrated in \cref{fig:AcOrch_pipeline}, leveraging its task orchestration strategy to achieve efficient overlap and scheduling both across and within devices. In the first stage, sampling tasks are distributed in parallel to AIV1 (batches 1, 2, 3) and CPU (batches 4, 5, 6), with sampling results asynchronously delivered to the NPU via a shared queue. Within the NPU, AIV2 is responsible for feature gathering (green blocks), and AIC for training computation (red blocks). \notecolor{Each stage is decoupled through asynchronous interaction, effectively masking the latency of communication and data preparation (sampling and feature gathering steps), thereby improving data transfer efficiency and reducing pipeline bubbles.}

The second stage further employs asynchronous queues and double buffering to fully overlap feature gathering on AIV2 and training computation on AIC, enabling efficient data and computation switching across batches and maximizing NPU bandwidth and compute utilization. As shown in \cref{fig:AcOrch_pipeline}, AIV1 and CPU process batch sampling in parallel, and AIV2 can immediately commence feature gathering for any batch upon its sampling completion, achieving pipelined parallelism between sampling and gathering. The AIC then schedules training tasks based on data arrival order, realizing end-to-end overlap of data and computation. This two-stage pipeline ensures sustained high parallelism for tasks both across devices and within the NPU, significantly improving system throughput and resource utilization. Experimental results (see Fig. 6) demonstrate that, on the Reddit with a 2-layer GraphSAGE, \system achieves optimal AIC utilization and training time, with improvements of 61.65\% and 2.78×, respectively, compared to MindSporeGL. This optimization strategy fully harnesses the parallel potential of CPU, AIV, and AIC, and serves as a key enabler for efficient and scalable large-scale heterogeneous GNN training systems.

\subsection{Graph aggregation operations remapping}

A typical GNN training task involves both graph aggregation operations and NN operations. In the training phase of MindSporeGL, a node-centric GNN framework, graph aggregation is deployed on the AIV, resulting in poor AIC utilization within the AI Cores and and a dominant computational overhead on the AIV. To address this limitation, we migrate the graph aggregation operations to the AIC, formalizing it as a sparse-matrix dense-matrix multiplication (SpMM) between the graph adjacency matrix (sparse matrix)  and the vertex feature matrix (dense matrix) on the AIC.
This enables the majority of the training process to be executed on the AIC, thereby freeing up the AIV resources for other steps.

\section{Evaluation}

\subsection{Experimental setup}
\paragraph{Environments}

All experiments are conducted on a server equipped with Huawei Kunpeng 920 CPU (192 cores and 768 GB main memory in total) and eight Huawei Ascend 910B1 NPUs (24 AI cores, 24 AICs, 48 AIVs, and 64 GB device memory). The system runs the EulerOS 2.8 operating system, with CANN v8.0 as the NPU runtime environment and MindSpore v2.3.1 as the deep learning framework.

\paragraph{Datasets}


\cref{tab:Dataset} summarizes six representative graphs datasets, including Reddit\cite{reddit}, Amazon\cite{orkut_amazon}, Wiki-Talk\cite{10.1145/1772690.1772756}, Products\cite{products}, Livejournal\cite{Livejournal}, and Orkut\cite{orkut_amazon}, spanning domains such as social networks, co-purchasing, and online forums, with node and edge scales ranging from hundreds of thousands to millions. For Wiki-Talk, Livejournal and Orkut, which lack intrinsic node attributes, we randomly generate node features and labels. 
All datasets are randomly divided into training, validation, and test sets using an 80\%/10\%/10\% ratio.

\begin{table} 
\centering
\setlength{\tabcolsep}{10pt} 
\caption{Dataset description. $|V|$, $|E|$ , $\#\mathbb{F}$ and $\#\mathbb{L}$ represent the number of vertices, edges, features and labels.}
\vspace{1em}
\label{tab:Dataset}
\begin{tabular}{l l l l l}
\hline
\textbf{Dataset} & \textbf{$|V|$} & \textbf{$|E|$} & \textbf{$\#\mathbb{F}$} & \textbf{$\#\mathbb{L}$} \\
\hline

Reddit \cite{reddit}        & 232.96K  & 114.61M  & 602 & 41  \\
Amazon \cite{orkut_amazon}        & 1.57M    & 264.34M  & 200 & 107 \\
Wiki-Talk \cite{10.1145/1772690.1772756}     & 2.4M     & 10.0M     & 600 & 60  \\
Products \cite{products}      & 2.4M     & 123M    & 100 & 47  \\
Livejournal \cite{Livejournal}   & 4.85M    & 138M   & 600 & 60  \\
Orkut \cite{orkut_amazon}         & 3.1M     & 234M     & 600 & 20  \\
\hline
\end{tabular}
\end{table}

\paragraph{GNN models and baselines}
We employ two representative GNN models, GraphSAGE\cite{reddit} and GCN\cite{gcn}, to evaluate \system. The training batch size for all models is set to 1024, the network depth is set to 2 layers, and the sampling fan-out is configured as [25, 10]. We use the state-of-the-art Ascend-native GNN training framework, MindSporeGL, as the baseline for comparative experiments. 

\begin{figure}
  \centering
  \vspace{-0.1in}
  \includegraphics[width=1\linewidth]{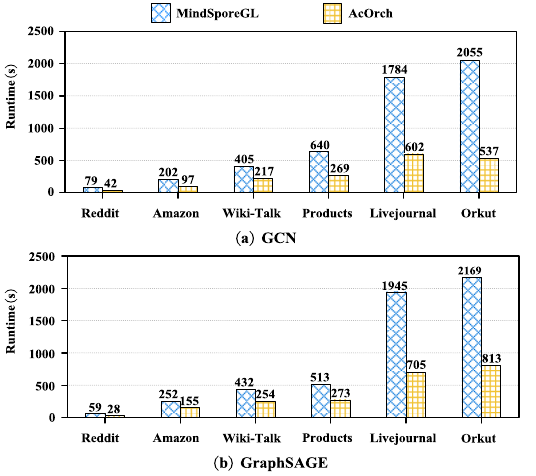}
   \vspace{-0.25in}
  \caption{The end-to-end training runtime of different systems.}
  \label{fig:shiyan1}
\end{figure}

\begin{figure}[t]
  \centering
  \includegraphics[width=1\linewidth]{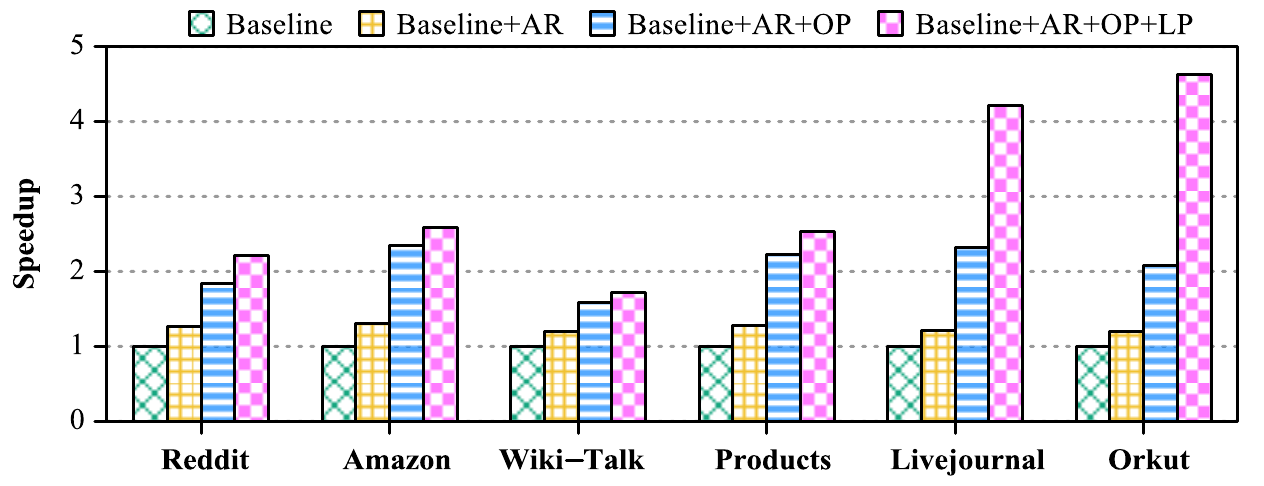}
  \caption{Performance gain analysis. "AR" indicates graph aggregation operations remapping, "OP" indicates orchestration and pipeline optimization, and "LP" indicates load-aware partitioning.}
  \label{fig:shiyan3}
\end{figure}

\subsection{Overall training performance comparison}


As shown in \cref{fig:shiyan1}, compared to MindSporeGL, \system achieves an average speedup of 2.31$\times$ across all datasets. As described in Section 3, MindSporeGL directly adopts a step-based task scheduling strategy, where sampling and feature collection during the data preparation stage are entirely bound to the CPU. This approach is constrained by CPU performance and PCIe bandwidth, resulting in prolonged NPU idleness. Furthermore, the training stage is dominated by graph aggregation operations executed on the AIV, leading to the coexistence of input bottlenecks (CPU) and computation bottlenecks (AIV), with very low AIC utilization. In contrast, \system employs finer-grained hardware resources such as AIC and AIV as scheduling units. By leveraging dual-path sampling during the data preparation stage, \system dynamically balances tasks between the CPU and AIV according to computational loads, while mapping graph aggregation operations to the AIC during training, thereby improving NPU utilization and achieving overall performance gains.
As the graph scale increases, \system delivers greater speedup, averaging 3.06$\times$ on datasets such as Livejournal and Orkut. This is because, at larger scales, the duration of each stage increases, limiting the pipeline overlap in MindSporeGL. In contrast, \system can fully exploit the advantages of workload balancing and pipeline overlap, effectively reducing hardware idle time and achieving more significant performance improvements. On other datasets (Reddit, Amazon, Wiki-Talk, and Products), \system still achieves an average speedup of 1.94$\times$. In summary, \system consistently demonstrates stable performance advantages across all datasets.


\subsection{Performance gain analysis of \system}



We analyze the performance improvements of graph aggregation operations remapping (AR), orchestration and pipeline optimization (OP), and load-aware partitioning (LP) on the GCN across six datasets. To ensure a fair comparison, we adopt a baseline based on the data-parallel implementation from the \system codebase, and sequentially integrate the three optimization methods one-by-one. The baseline method employs step-based task orchestration, where graph sampling is performed on the CPU, while the gathering and training stages are executed on the NPU. Pipeline optimization is not applied, and graph aggregation during training utilizes the AIV operator. Figure \cref{fig:shiyan3} presents the normalized speedup results.
Compared to the baseline, AR accelerates the training phase by mapping the graph aggregation operation onto the AIC, yielding a 20\% to 31\% improvement in performance.
Based on baseline+AR, OP further splits the sampling tasks and introduces a two-stage pipeline, resulting in a 39\% to 111\% increase in performance. In extremely sparse, mini-batch sampling scenarios such as Wiki-Talk, the fine granularity of sampling tasks increases the proportion of system overhead from process switching and batch scheduling, with only a 39\% acceleration achieved. On top of baseline+AR+OP, LP applies load-aware task partitioning and scheduling, achieving an average performance improvement of 97\% and further balancing the sampling workload across heterogeneous resources. For datasets such as Livejournal and Orkut, LP achieves speedups of 1.89$\times$ and 2.55$\times$, respectively, mainly because static task partitioning tends to concentrate the load and create bottlenecks in large-scale, power-law distributed graphs, whereas LP effectively balances the load and smooths the pipeline execution.

\begin{figure}[t]
  \centering
  \includegraphics[width=1\linewidth]{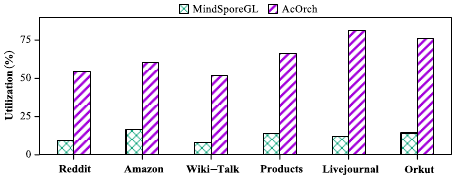}
  \caption{Comparison of AIC utilization of different systems.}
  \label{fig:shiyan2}
\end{figure}

\begin{figure}[t]
  \centering
  \includegraphics[width=1\linewidth]{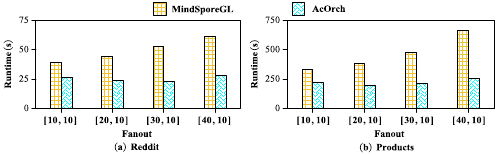}
  \caption{Per-epoch runtime of different systems with different fanout.}
  \label{fig:shiyan-fanout}
\end{figure}

\begin{figure}[t]
  \centering
  \includegraphics[width=0.95\linewidth]{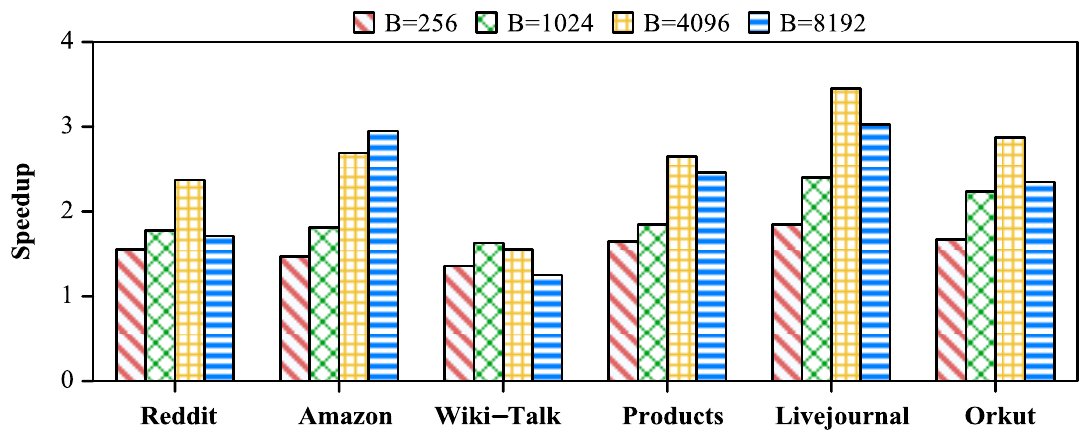}
  \caption{The speedup compared to MindSporeGL with different batch sizes (B).}
  \label{fig:shiyan4}
\end{figure}

\subsection{AIC utilization}


AIC accounts for more than 90\% of the computing power of the Ascend AI processor, so we use AIC utilization as a proxy for overall NPU utilization. Using the \texttt{msprof} profiling tool, we evaluate AIC utilization during GCN training on six datasets for both \system and MindSporeGL. As shown in Figure~\ref{fig:shiyan2}, \system achieves a significantly higher average AIC utilization of 52.63\%, whereas MindSporeGL attains only 12.29\%. 
The low AIC utilization in MindSporeGL is primarily due to pipeline stalls caused by CPU-based data preparation, resulting in prolonged NPU idleness. Additionally, most NPU computation is concentrated on graph aggregation operators running on the AIV. In contrast, \system employs a compute-aware dual-path sampling strategy during data preparation to achieve load balancing and utilizes shared queues and a two-stage pipeline to efficiently overlap heterogeneous computing resources. 

\subsection{Sensitivity study}

\paragraph{Performance with varying fanout}


In this experiment, we compare the performance of \system and MindSporeGL when training GraphSAGE with different fanout settings on the Reddit and Products datasets. As shown in \cref{fig:shiyan-fanout}, when the fanout is [10,10], \system achieves an average speedup of 1.49$\times$ over MindSporeGL. For fanout settings of [20,10], [30,10], and [40,10], the average speedups reach 1.89$\times$, 2.28$\times$, and 2.39$\times$, respectively. The setting of fanout will not affect the effectiveness of \system. Furthermore, as fanout increases, the sampling and feature collection bottlenecks of MindSporeGL become more severe. In contrast, \system leverages a computation-aware heterogeneous sampling partitioning mechanism, which maintains efficient overlap across all stages even as the sampling workload increases, resulting in a consistent improvement in speedup as fanout grows.

\paragraph{Performance with varying batch sizes}



To evaluate the effectiveness of \system under different batch sizes, we train GraphSAGE with varying batch sizes on the six datasets. As shown in \cref{fig:shiyan4}, when the batch size is 256, \system achieves an average speedup of 1.64$\times$ compared to MindSporeGL. When the batch sizes are 1024, 4096, and 8192, the average speedups reach 2.12$\times$, 2.83$\times$, and 2.70$\times$, respectively. 
Overall, the speedup increases with batch size at first and then decreases. When the batch size is small, the task load is light and the effect of heterogeneous scheduling is limited. 
With a larger batch size, the scheduling mechanism of \system effectively improves the parallelism of heterogeneous resources, resulting in improved performance acceleration.


\paragraph{Performance with varying workload partition ratios}

\begin{figure}[t]
  \centering
  \includegraphics[width=1\linewidth]{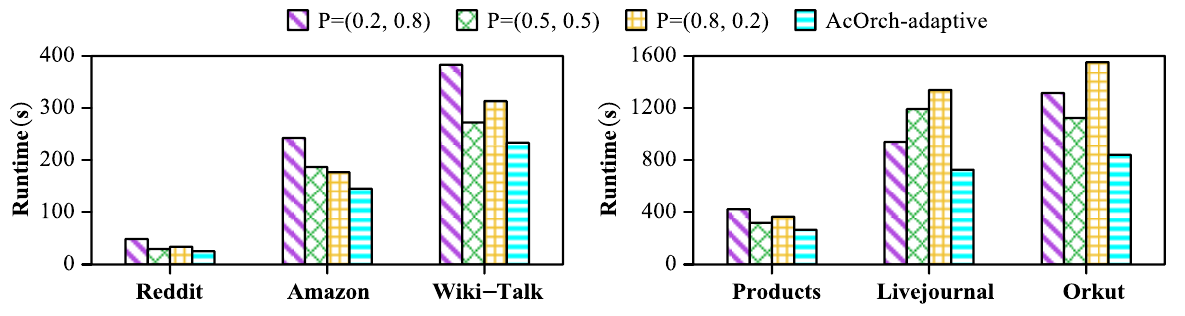}
  \caption{Per-epoch runtime of different systems with different workload partition ratios. P=(p, 1–p) denotes the AIV/CPU sampling node partition ratio, and the \system-adaptive indicates computation-aware workload partition.}
  \label{fig:shiyan-workload}
\end{figure}


In this experiment, we evaluate the efficiency of \system's computation-aware adaptive partitioning strategy against three fixed sampling workload partition ratios using GraphSAGE on six datasets. As shown in \cref{fig:shiyan-workload}, the workload partition ratio significantly affects training efficiency, with an average speedup of up to 1.75$\times$ between the best and worst fixed partitioning schemes. In contrast, \system's adaptive scheduling consistently outperforms all fixed strategies and achieves speedups ranging from 1.2x to 1.3x over the best-fixed partition ratio. This improvement is primarily due to its ability to dynamically adjust the partition ratio based on task load, resulting in better load balancing.



\begin{figure}[t]
  \centering
  \includegraphics[width=1\linewidth]{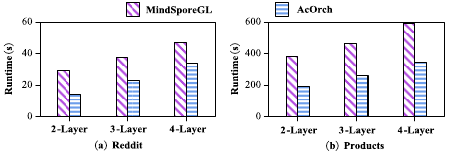}
  \caption{\notecolor{Per-epoch runtime of different systems with different model depths.}}
  \label{fig:shiyan-depth}
\end{figure}

\paragraph{\notecolor{Performance with varying model depths}}
\notecolor{In this experiment, we compare the performance of \system and MindSporeGL when training GraphSAGE with different model depths on the Reddit and Products datasets. As shown in \cref{fig:shiyan-depth}, the size of sampled subgraphs grows exponentially as the number of GNN layers increases, resulting in a significant increase in training time for both \system and MindSporeGL. Nevertheless, \system consistently outperforms MindSporeGL across 2-layer, 3-layer, and 4-layer GraphSAGE models, achieving average speedups of 1.90×, 1.76×, and 1.47×, respectively, demonstrating that the performance advantages of \system’s task scheduling and pipelined optimization strategies are maintained with increasing model depth.}

\subsection{Accuracy comparison}
\begin{figure}[t]
  \centering
  \includegraphics[width=1\linewidth]{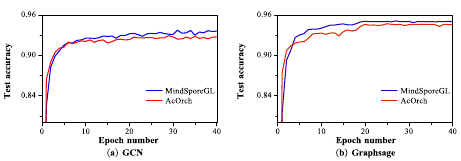}
  \caption{Epoch-to-accuracy.}
  \label{fig:shiyan5}
\end{figure}

We conduct epoch-accuracy comparisons for GCN and GraphSAGE models across different systems, as shown in \cref{fig:shiyan5}.For both GCN and GraphSAGE, \system consistently achieves test accuracy comparable to that of MindSporeGL throughout the entire training process.
As the number of epochs increases, the accuracy of both systems improves rapidly and converges after 40 epochs, ultimately reaching nearly identical accuracy levels.

\begin{table}[!h]
  \renewcommand{\arraystretch}{0.9}
  \centering
  \caption{{\notecolor{Runtime breakdown (in seconds and percentage of total time) for 50 epochs on six datasets with GraphSAGE.}}}
  \vspace{0.1in}
  \begin{tabular}{l ccc}
    \toprule
    Dataset & Partition           & Train               & Total  \\
    \midrule
    Reddit      & 0.52 / 2.93\%   & 17.09 / 96.42\%  & 17.72 \\
    Amazon      & 3.09 / 3.34\%   & 88.97 / 96.04\%  & 92.64 \\
    Wiki-Talk   & 3.82 / 1.97\%   & 188.63 / 97.08\% & 194.32 \\
    Products    & 9.72 / 4.50\%   & 204.78 / 94.71\% & 216.21 \\
    Livejournal & 35.52 / 5.69\%  & 582.61 / 93.39\% & 623.85 \\
    Orkut       & 22.14 / 3.91\%  & 539.95 / 95.08\% & 566.15 \\
    \bottomrule
  \end{tabular}
  \label{tab:partition-breakdown}
\end{table}

\subsection{preprocessing overhead analysis}


\notecolor{
To evaluate the additional overhead introduced by AcOrch's partitioning optimization, we measure the runtime cost of the partitioning operation for each mini-batch as well as its reuse efficiency during training. The corresponding results are presented in \cref{tab:partition-breakdown}. Specifically, workload partitioning is performed for every mini-batch and requires only sorting and linear assignment of node workloads, resulting in a relatively low algorithmic complexity ($O(V\log V)$, where $V$ denotes the number of nodes in the mini-batch). The partitioning results are reused via the HBM and repartitioning is triggered only when the performance metric reaches a predefined threshold, which effectively amortizes the partitioning overhead. On average, the partition step accounts for only 3.7\% of the total runtime in 50-epoch GraphSAGE training.
}

\begin{table}[!h]
  \renewcommand{\arraystretch}{0.9}
  \centering
  \caption{{\notecolor{Tail latency impact on throughput for dual-path sampling.}}}
  \vspace{0.1in}
  \begin{tabular}{l ccc}
    \toprule
    Dataset & Batch latency & P99 latency & Throughput degradation \\
    \midrule
    Reddit   & 38.8 ms & 40.3 ms & 983 requests/s (3.7\%)  \\
    Products & 52.3 ms & 55.1 ms & 993 requests/s (5.1\%)  \\ 
    \bottomrule
  \end{tabular}
  \label{tab:tail_latency}
\end{table}

\subsection{Impact of tail latency on throughput}


\notecolor{
To evaluate the impact of CPU and AIV load imbalance in dual-path sampling on \system training throughput in the worst case, we measure the end-to-end execution time of 1,000 batches (each of size 1024) on the Reddit and Products datasets, and compute the average latency per batch (Batch latency), the 99th percentile latency (P99 latency), and the throughput degradation. \cref{tab:tail_latency} shows that throughput decreases by 983 requests/s (3.7\%) for Reddit and 993 requests/s (5.1\%) for Products, respectively. P99 latency is close to the average latency, indicating that the overall system performance remains stable under tail latency conditions.
}

\begin{figure}[t]
  \centering
  \includegraphics[width=0.8\linewidth]{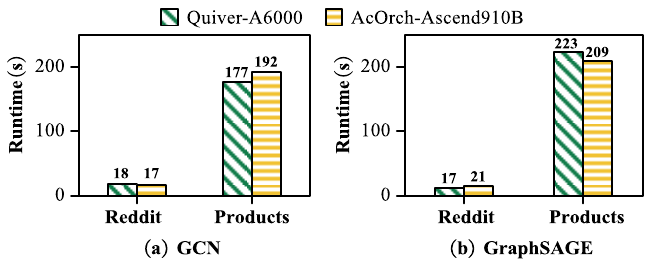}
  \caption{\notecolor{The end-to-end runtime compared to the GPU-based GNN training framework.}}
  \label{fig:shiyan-gpu}
\end{figure}

\begin{figure}[t]
  \centering
  \includegraphics[width=1\linewidth]{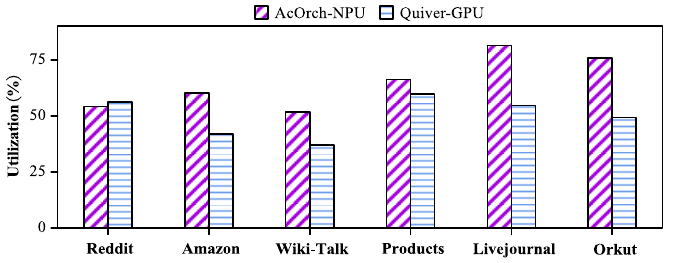}
  \caption{Comparison of resource utilization of different systems.}
  \label{fig:shiyan-uti}
\end{figure}

\begin{figure}[t]
  \centering
  \includegraphics[width=1\linewidth]{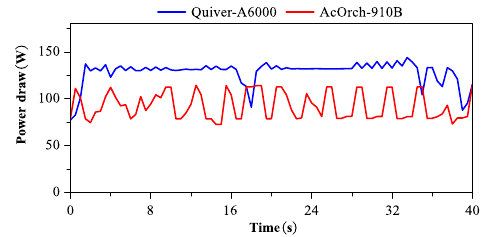}
  \caption{\notecolor{Power draw (in watts) sampled every 500ms over a 40-second training window.}}
  \label{fig:shiyan-power}
\end{figure}

\subsection{Comparisons with GPU-based GNN training framework}

\notecolor{In this subsection, we evaluate the performance of \system in comparison with Quiver [53], a representative GPU-optimized GNN training framework. Experiments are conducted on the GCN and GraphSAGE models, with Quiver evaluated on NVIDIA A6000 GPU, and \system evaluated on Ascend 910B. We report metrics including end-to-end training performance, compute resource utilization, and power consumption.}

\paragraph{Performance Comparison}
\notecolor{We evaluated \system on Reddit and Products using GCN and GraphSAGE, comparing against Quiver on NVIDIA A6000, as shown in \cref{fig:shiyan-gpu}. On average, \system on Ascend 910B achieves 95.1\% of Quiver’s end-to-end performance, indicating that NPU-based training can achieve performance close to a GPU-optimized framework.}

\paragraph{Resource Utilization}
\notecolor{We evaluate compute unit utilization using \texttt{msprof} on Ascend 910B and nvidia-smi on the GPU. As shown in \cref{fig:shiyan-uti}, \system exhibits higher compute resource utilization, with an average improvement of 15.11\% compared to Quiver. The gain is mainly attributed to \system’s sample-driven task orchestration and fine-grained pipeline optimization, which help keep the heterogeneous compute units of the Ascend 910B more consistently busy.}

\paragraph{Power Consumption}
\notecolor{We measure energy efficiency by recording power draw (in watts) every 500 milliseconds over a 40-second interval during model training. On Ascend 910B, power is monitored using \texttt{npu-smi}, while on NVIDIA GPUs, \texttt{nvidia-smi} is used. \cref{fig:shiyan-power} shows the results. \system on Ascend 910B has an average power draw of 91.85 W, with a peak of 115.6 W. In comparison, Quiver on A6000 averages 109.71 W, with a peak of 144.1 W. The results indicate that \system can operate with lower power consumption while maintaining comparable training performance.}

\subsection{Memory pressure and bandwidth utilization analysis}

\begin{table*}[h]
\centering
  \captionsetup{justification=centering} 
  \caption{\notecolor{Data interconnect bandwidth and utilization during sampling}}
\begin{tabular}{lcccc}
\toprule
Metric & Average bandwidth & Peak bandwidth  & Average utilization & Peak utilization \\
\midrule
PCIe 4.0 & 40.7 GB/s   & 54.8 GB/s    & 63.6\%  & 85.7\%  \\
MTE & 0.87 TB/s   &    1.1 TB/s    & 70.9\%  & 89.5\%  \\
\bottomrule
\label{tab:bandwidth}
\end{tabular}
\end{table*}

\notecolor{\cref{tab:bandwidth} presents the bandwidth and utilization metrics for PCIe 4.0 (host-to-device interconnect) and MTE (on-chip data mover between HBM and AI Core units) during the sampling stage on the Products dataset with the GraphSAGE model. MTE achieves an average bandwidth of 0.87 TB/s and a peak bandwidth of 1.1 TB/s, with average utilization at 70.9\% and peak utilization at 89.5\%. In comparison, PCIe 4.0 records an average bandwidth of 40.7 GB/s and a peak bandwidth of 54.8 GB/s, with average utilization at 63.6\% and peak utilization at 85.7\%. Overall, MTE demonstrates higher bandwidth utilization than PCIe, indicating that intra-core data movement is more efficient than interconnect data transfer under the current design. However, MTE utilization does not reach the hardware limit, suggesting further potential for optimization in memory hierarchy and data scheduling.
In future work, we further leverage the multi-level cache architecture of NPUs, such as L0/L1/L2 buffers and HBM, to implement hierarchical caching and dynamic prefetching not only for node features but also for graph topology. This approach reduces memory access overhead, improves cache hit rate and MTE bandwidth utilization, and optimizes overall system performance for large-scale graph sampling and training. }

\section{Discussion}

\begin{table*}[!t]
  \centering
  \captionsetup{justification=centering} 
  \caption{\notecolor{Support for \system's techniques across other NPU architectures.}}
  \label{tab:npu-generalization}
  \begin{tabular}{lccc}
    \toprule
    \textbf{Technique} & \textbf{Sample-driven task orchestration} & 
   \textbf{Fine-grained pipeline optimization} \\
    \cmidrule(r){2-2} \cmidrule(l){3-3}
    \textbf{Hardware feature} & \textbf{Heterogeneous compute units} & \textbf{Multi-level memory hierarchies}  \\
    \midrule
    Ascend 910B [33] & \checkmark~(AIC/AIV) & \checkmark~(L1, L2, DDR)  \\
    TPU v4 [35]      & \checkmark~(MXU/VPU) & \checkmark~(Unified buffer, HBM) \\
    Gaudi2 [36]      & \checkmark~(MME/Tensor Core) & \checkmark~(SRAM, DRAM) \\
    MLU370 [55]      & \checkmark~(Tensor/Vector Function Unit) & \checkmark~(Local Buffer, GDDR) \\
    \bottomrule
  \end{tabular}
\end{table*}

\subsection{Extension to other accelerator architecturess}

\paragraph{Applicability to other NPUs}
\notecolor{Although \system is implemented and evaluated on the Ascend 910B platform, its design principles can be generalized to other NPUs with comparable compute and memory architectures. As shown in \cref{tab:npu-generalization}, our approach relies on two key architectural features: 
(1) \textbf{Heterogeneous compute units}, such as AIC/AIV on Ascend, MXU/VPU on TPU v4, MME/Tensor processor core on Gaudi2, and Tensor/Vector function units on MLU370. These units execute matrix and vector operations independently within or across cores, enabling different pipeline stages to proceed in parallel; 
(2) \textbf{Multi-level memory hierarchies}, including L1/L2/DDR on Ascend, unified buffer with HBM on TPU v4, and on-chip SRAM/DRAM or local buffer/GDDR on Gaudi2 and MLU370. These hierarchies sustain data supply to compute units and allow explicit buffer management, thereby reducing idle cycles and supporting fine-grained pipeline execution. In summary, the combination of heterogeneous compute resources and programmable memory hierarchies allows \system’s sample-driven task orchestration and pipeline optimization to be effectively adapted to other NPU backends.}

\paragraph{Limitations on GPUs and FPGAs}

\notecolor{Applying \system to GPUs or FPGAs presents challenges due to differences in architecture and execution models. The programming model of GPUs is based on SIMT execution with warp-level scheduling, where latency is hidden by massive multithreading rather than explicit pipeline control. This mechanism limits fine-grained intra-core pipelining and buffer management at the granularity required by \system. FPGAs, while supporting customized pipelines, rely on hardware-level design through RTL or HLS. Their execution model requires static compilation and lacks the flexibility of software-level task orchestration that \system exploits on NPUs. Therefore, although the principle of sample-driven task orchestration remains broadly relevant, the proposed techniques are most naturally extended to NPUs, where architectural features directly support the necessary heterogeneity and buffer-level programmability.}

\subsection{Limitations and future work}



\notecolor{Currently, \system operates under the assumption that all data can be stored in CPU memory, which limits the maximum graph size that can be processed by the available CPU memory. To address this limitation, we propose two directions for future research.}

\paragraph{Distributed training extension} 
\notecolor{Distributed \system represents a promising direction for improving scalability. Drawing on existing distributed graph neural network systems such as NeutronStar\cite{neutronstar} and DGL\cite{dgl}, we plan to extend NPU-based distributed systems. Specifically, \system can be deployed on each node, fully utilizing heterogeneous resources to accelerate single-node training. This process requires efficient feature synchronization across NPUs and dynamic task scheduling among nodes.}

\paragraph{Feature management based on NPU multi-level buffer} 
\notecolor{By leveraging the multi-level buffer of NPUs, including L0/L1/L2 buffer, and HBM, and accounting for node activity and access locality, we aim to design hierarchical feature caching and dynamic prefetching mechanisms. This will improve cache hit rates during the gathering stage and enable efficient memory management for large-scale graph GNN sampling and training scenarios.}
\section{Related work}
\paragraph{Sampling-based GNN training}

In large-scale GNN training, GraphSAGE first introduces the sampling and message aggregation paradigm. In CPU-GPU heterogeneous environments, existing studies primarily adopt step-based orchestration to alleviate sampling and gathering bottlenecks. Systems such as TurboGNN\cite{turboGNN} and Nextdoor\cite{nextdoor_eurosys21} utilize GPUs to accelerate sampling, while Quiver\cite{Quiver_corr_2023} and PaGraph\cite{pagraph} leverage GPUs for gathering and propose various caching strategies\cite{gnnfeaturechache-22}. NeutronOrch\cite{neuorch}, as one of the current state-of-the-art systems for sampling-based GNN training, introduces a layer-wise decoupled scheduling approach, assigning computation tasks of different layers to CPUs and GPUs. Some works explore the use of FPGAs to accelerate GNN operations in CPU-FPGA settings\cite{cpu-fpga}. The layer-wise decoupled scheduling of NeutronOrch relies on the logically unified resources and flexible scheduling capabilities of GPUs. 
However, on CPU-NPU platforms, NPUs such as Ascend AI processors exhibit architectural heterogeneity and coarse-grained kernels, requiring static manual scheduling. This increases orchestration complexity, making it difficult for existing methods to be ported to NPU platforms.

\paragraph{Research progress on Ascend AI processors}


The Huawei Ascend AI processor, as a new-generation NPUs, efficiently supports mainstream DNN computations and is compatible with deep learning frameworks such as PyTorch, TensorFlow, and native MindSpore\cite{ascend_scalable}. Benchmark evaluations like AIbench\cite{Ascend_bench_4} indicate that Ascend AI processors can achieve computational power comparable to or surpassing contemporary GPUs across various DNN scenarios, with performance on MindSpore exceeding that on PyTorch\cite{ascend_scalable,Ascend_bench_1,Ascend_bench_2}. 
NeutronAscend \cite{ai2025neutronascend} is a GNN training framework for Ascend NPUs that jointly optimizes inter–AI Core parallelism and intra-core kernel execution to accelerate large-scale GNN training.
In addition, application-level advancements have been achieved on Ascend AI processors. The Verrocchio\cite{ascend_Verrocchio_jpdc_23} performance model can accurately predict operator latency, providing theoretical guidance for operator scheduling and system optimization. HPL-MxP\cite{ascend-cs24} introduces a mixed-precision LU decomposition algorithm leveraging NPU-CPU collaboration, achieving performance improvement through multi-level heterogeneous pipelining and matrix multiplication fusion. Cube-fx\cite{ascend-martrix} maps Taylor expansion onto matrix multiply-accumulate units, achieving performance improvement over the vectorized Horner’s method on Ascend AI processors.


\section{Conclusion}

This work addresses the problem of low resource utilization in sampling-based GNN training on CPU-NPU heterogeneous platforms. We propose an efficient task orchestration mechanism and a two-level pipeline mechanism tailored to NPU heterogeneous compute units, and implement them in the \system system. By leveraging fine-grained task partitioning and coordinated use of heterogeneous resources, \system significantly improves both training throughput and hardware utilization. Experiments on the Ascend 910B show that \system achieves an average speedup of 2.31$\times$ over the mainstream framework MindSporeGL across multiple real-world graph datasets, along with an average of 52.63\% improvement in AIC utilization, fully unlocking the performance potential of the NPU. Future work will explore further optimization for GNN training in multi-NPU heterogeneous environments, which necessitate more complex communication-computation orchestration.


\section*{Acknowledgements}
This work was supported by the National Key Research and Development Program of China (No. 2023YFB4503601), the National Natural Science Foundation of China (U2241212, 62461146205), the Distinguished Youth Foundation of Liaoning Province (No. 2024021148-JH3/501).

\bibliographystyle{fcs}
\bibliography{sample-base}

\begin{thebibliography}{10}

\bibitem{neuorch}
Ai~X, Wang Q, Cao C, Zhang Y, Chen C, Yuan H, Gu~Y, Yu~G.
\newblock {NeutronOrch:} rethinking sample-based {GNN} training under {CPU-GPU} heterogeneous environments.
\newblock Proceedings of the VLDB Endowment, 2024, 17(8): 1995--2008

\bibitem{gnnworkssocial}
Mu~N, Zha D, He~Y, Tang Z.
\newblock Graph attention networks for neural social recommendation.
\newblock In: Proceedings of the 31st {IEEE} International Conference on Tools with Artificial Intelligence.
\newblock 2019,  1320--1327

\bibitem{neutronstar}
Wang Q, Zhang Y, Wang H, Chen C, Zhang X, Yu~G.
\newblock {NeutronStar:} distributed {GNN} training with hybrid dependency management.
\newblock In: Proceedings of 2022 International Conference on Management of Data.
\newblock 2022,  1301--1315

\bibitem{dgl}
Wang M, Yu~L, Zheng D, Gan Q, Gai Y, Ye~Z, Li~M, Zhou J, Huang Q, Ma~C, Huang Z, Guo Q, Zhang H, Lin H, Zhao J, Li~J, Smola A~J, Zhang Z.
\newblock Deep graph library: towards efficient and scalable deep learning on graphs.
\newblock 2019,  arXiv preprint arXiv: 1909.01315

\bibitem{XGCN}
Song X, Huang H, Lian J, Jin H.
\newblock {XGCN:} a library for large-scale graph neural network recommendations.
\newblock Frontiers of Computer Science, 2024, 18(3): 183343

\bibitem{gcn}
Kipf T~N, Welling M.
\newblock Semi-supervised classification with graph convolutional networks.
\newblock In: Proceedings of the 5th International Conference on Learning Representations.
\newblock 2017

\bibitem{turboGNN}
Wu~W, Shi X, He~L, Jin H.
\newblock {TurboGNN:} improving the end-to-end performance for sampling-based {GNN} training on {GPUs}.
\newblock IEEE Transactions on Computers, 2023, 72(9): 2571--2584

\bibitem{gnnspmethod_ieeejas_2022}
Liu X, Yan M, Deng L, Li~G, Ye~X, Fan D.
\newblock Sampling methods for efficient training of graph convolutional networks: {a} survey.
\newblock IEEE/CAA Journal of Automatica Sinica, 2022, 9(2): 205--234

\bibitem{gnnsurvey_tnn21}
Wu~Z, Pan S, Chen F, Long G, Zhang C, Yu~P~S.
\newblock A comprehensive survey on graph neural networks.
\newblock IEEE Transactions on Neural Networks and Learning Systems, 2021, 32(1): 4--24

\bibitem{fcs-gnnsurvey}
Liu J, Chen S, Shen L.
\newblock A comprehensive survey on graph neural network accelerators.
\newblock Frontiers of Computer Science, 2025, 19(2): 192104

\bibitem{fcs-dynamicgnnsurvey}
Zheng Y, Yi~L, Wei Z.
\newblock A survey of dynamic graph neural networks.
\newblock Frontiers of Computer Science, 2025, 19(6): 196323

\bibitem{sancusgnn_ijcai_2023}
Peng J, Chen Z, Shao Y, Shen Y, Chen L, Cao J.
\newblock {SANCUS:} staleness-aware communication-avoiding full-graph decentralized training in large-scale graph neural networks.
\newblock Proceedings of the VLDB Endowment, 2022, 15(9): 1937--1950

\bibitem{gnnspcheat-tcss24}
Tian Y, Liu G, Wang J, Zhou M.
\newblock {ASA-GNN:} adaptive sampling and aggregation-based graph neural network for transaction fraud detection.
\newblock IEEE Transactions on Computational Social Systems, 2024, 11(3): 3536--3549

\bibitem{gnnsphetero-mbec24}
Mishra S, Singh G, Bhattacharya M.
\newblock Tissue specific tumor-gene link prediction through sampling based {GNN} using a heterogeneous network.
\newblock Medical {\&} Biological Engineering {\&} Computing, 2024, 62(8): 2499--2510

\bibitem{gnnspfuse-cluster24}
Wei X, Tang W, Qi~H, Yue H.
\newblock {PGSampler:} accelerating {GPU}-based graph sampling in {GNN} systems via workload fusion.
\newblock In: Proceedings of 2024 {IEEE} International Conference on Cluster Computing.
\newblock 2024,  51--61

\bibitem{gnnspcase_sigops_2021}
Serafini M.
\newblock Scalable graph neural network training{:} the case for sampling.
\newblock ACM SIGOPS Operating Systems Review, 2021, 55(1): 68--76

\bibitem{EfficientSampling_train}
Bai Y, Li~C, Lin Z, Wu~Y, Miao Y, Liu Y, Xu~Y.
\newblock Efficient data loader for fast sampling-based {GNN} training on large graphs.
\newblock IEEE Transactions on Parallel and Distributed Systems, 2021, 32(10): 2541--2556

\bibitem{Zhu_2024}
Zhu Z, Wang P, Hu~Q, Li~G, Liang X, Cheng J.
\newblock {FastGL:} a {GPU}-efficient framework for accelerating sampling-based {GNN} training at large scale.
\newblock In: Proceedings of the 29th {ACM} International Conference on Architectural Support for Programming Languages and Operating Systems.
\newblock 2024,  94--110

\bibitem{alimohammadi2023localgraphlimitsperspective}
Alimohammadi Y, Ruiz L, Saberi A.
\newblock A local graph limits perspective on sampling-based {GNNs}.
\newblock 2023,  arXiv preprint arXiv: 2310.10953

\bibitem{10.1145/3492321.3519557}
Yang J, Tang D, Song X, Wang L, Yin Q, Chen R, Yu~W, Zhou J.
\newblock {GNNLab:} a factored system for sample-based {GNN} training over {GPUs}.
\newblock In: Proceedings of the 17th European Conference on Computer Systems.
\newblock 2022,  417--434

\bibitem{10.1145/3534678.3539038}
Min S, Wu~K, Hidayetoglu M, Xiong J, Song X, Hwu W.
\newblock Graph neural network training and data tiering.
\newblock In: Proceedings of the 28th {ACM} {SIGKDD} Conference on Knowledge Discovery and Data Mining.
\newblock 2022,  3555--3565

\bibitem{10.1145/3589311}
Zhang X, Shen Y, Shao Y, Chen L.
\newblock {DUCATI:} a {Dual-Cache} training system for graph neural networks on giant graphs with the {GPU}.
\newblock Proceedings of the ACM on Management of Data, 2023, 1(2): 166:1--166:24

\bibitem{10.14778/3476249.3476264}
Min S, Wu~K, Huang S, Hidayetoglu M, Xiong J, Ebrahimi E, Chen D, Hwu W~W.
\newblock Large graph convolutional network training with {GPU}-oriented data communication architecture.
\newblock Proceedings of the VLDB Endowment, 2021, 14(11): 2087--2100

\bibitem{chen2018fastgcnfastlearninggraph}
Chen J, Ma~T, Xiao C.
\newblock {FastGCN:} fast learning with graph convolutional networks via importance sampling.
\newblock In: Proceedings of the 6th International Conference on Learning Representations.
\newblock 2018

\bibitem{cpu-fpga}
Zhang B, Kuppannagari S~R, Kannan R, Prasanna V~K.
\newblock Efficient neighbor-sampling-based {GNN} training on {CPU-FPGA} heterogeneous platform.
\newblock In: Proceedings of 2021 {IEEE} High Performance Extreme Computing Conference, {HPEC} 2021.
\newblock 2021,  1--7

\bibitem{huang2024freshgnnreducingmemoryaccess}
Huang K, Jiang H, Wang M, Xiao G, Wipf D, Song X, Gan Q, Huang Z, Zhai J, Zhang Z.
\newblock {FreshGNN:} reducing memory access via stable historical embeddings for graph neural network training.
\newblock Proceedings of the VLDB Endowment, 2024, 17(6): 1473--1486

\bibitem{fey2021gnnautoscalescalableexpressivegraph}
Fey M, Lenssen J~E, Weichert F, Leskovec J.
\newblock {GNNAutoScale}: scalable and expressive graph neural networks via historical embeddings.
\newblock In: Proceedings of the 38th International Conference on Machine Learning.
\newblock 2021,  3294--3304

\bibitem{mohoney2021mariuslearningmassivegraph}
Mohoney J, Waleffe R, Xu~H, Rekatsinas T, Venkataraman S.
\newblock {Marius:} learning massive graph embeddings on a single machine.
\newblock In: Proceedings of the 15th {USENIX} Symposium on Operating Systems Design and Implementation.
\newblock 2021,  533--549

\bibitem{WholeGraph-sc22}
Yang D, Liu J, Qi~J, Lai J.
\newblock {WholeGraph:} a fast graph neural network training framework with multi-{GPU} distributed shared memory architecture.
\newblock In: Proceedings of International Conference for High Performance Computing, Networking, Storage and Analysis.
\newblock 2022,  54:1--54:14

\bibitem{sagnn-iscas23}
Wang H, Zhang S, Feng K, Wang M, Yang Z.
\newblock {SaGNN:} a sample-based {GNN} training and inference hardware accelerator.
\newblock In: Proceedings of 2023 {IEEE} International Symposium on Circuits and Systems.
\newblock 2023,  1--5

\bibitem{gnnsplossspeed-ispa24}
Song Y, Wang Y, Xiong C, Wang T, Tang P.
\newblock An efficient sampling-based {SpMM} kernel for balancing accuracy and speed in {GNN} inference.
\newblock In: Proceedings of 2024 {IEEE} International Symposium on Parallel and Distributed Processing with Applications.
\newblock 2024,  468--475

\bibitem{gnnfeaturechache-22}
He~Y, Lai Z, Ran Z, Zhang L, Li~D.
\newblock {SCGraph:} accelerating sample-based {GNN} training by staged caching of features on {GPUs}.
\newblock In: Proceedings of 2022 {IEEE} International Conference on Parallel {\&} Distributed Processing with Applications, Big Data {\&} Cloud Computing, Sustainable Computing {\&} Communications, Social Computing {\&} Networking.
\newblock 2022,  106--113

\bibitem{ascend_scalable}
Liao H, Tu~J, Xia J, Liu H, Zhou X, Yuan H, Hu~Y.
\newblock {Ascend:} a scalable and unified architecture for ubiquitous deep neural network computing {:} industry track paper.
\newblock In: Proceedings of 2021 IEEE International Symposium on High-Performance Computer Architecture.
\newblock 2021,  789--801

\bibitem{Ascend_bench_2}
Zhu Z, Wang B, Yang C, Zhu R, Zhou M, Zheng N.
\newblock Performance evaluation of {MindSpore} and {PyTorch} based on {Ascend} {NPU}.
\newblock In: Proceedings of the 29th {IEEE} International Conference on Parallel and Distributed Systems.
\newblock 2023,  1826--1832

\bibitem{TPUv4}
Jouppi N~P, Young C, Patil N, Patterson D~A, Agrawal G, Bajwa R, Bates S, Bhatia S, Boden N, Borchers A, Boyle R, Cantin P, Chao C, Clark C, Coriell J, Daley M, Dau M, Dean J, Gelb B, Ghaemmaghami T~V, Gottipati R, Gulland W, Hagmann R, Ho~C~R, Hogberg D, Hu~J, Hundt R, Hurt D, Ibarz J, Jaffey A, Jaworski A, Kaplan A, Khaitan H, Killebrew D, Koch A, Kumar N, Lacy S, Laudon J, Law J, Le~D, Leary C, Liu Z, Lucke K, Lundin A, MacKean G, Maggiore A, Mahony M, Miller K, Nagarajan R, Narayanaswami R, Ni~R, Nix K, Norrie T, Omernick M, Penukonda N, Phelps A, Ross J, Ross M, Salek A, Samadiani E, Severn C, Sizikov G, Snelham M, Souter J, Steinberg D, Swing A, Tan M, Thorson G, Tian B, Toma H, Tuttle E, Vasudevan V, Walter R, Wang W, Wilcox E, Yoon D~H.
\newblock In-datacenter performance analysis of a tensor processing unit.
\newblock In: Proceedings of the 44th Annual International Symposium on Computer Architecture, {ISCA} 2017.
\newblock 2017,  1--12

\bibitem{Gaudi2}
Medina E, Dagan E.
\newblock Habana labs purpose-built {AI} inference and training processor architectures: Scaling {AI} training systems using standard ethernet with gaudi processor.
\newblock {IEEE} Micro, 2020, 40(2): 17--24

\bibitem{Ascend_bench_4}
Xiao Y, Wang Z.
\newblock {AIbench:} a tool for benchmarking {Huawei} {Ascend} {AI} processors.
\newblock CCF Transactions on High Performance Computing, 2024, 6(2): 115--129

\bibitem{dhar2024ascendcc}
Dhar A, Thorens C, Lazier L~M, Cavigelli L.
\newblock {Ascend-CC:} confidential computing on heterogeneous {NPU} for emerging generative {AI} workloads.
\newblock 2024, arXiv preprint arXiv: 2407.11888

\bibitem{tpu_corr25}
Wongpanich A, Oguntebi T, Paredes J~B, Wang Y~E, Phothilimthana P~M, Mitra R, Zhou Z, Kumar N, Reddi V~J.
\newblock Machine learning fleet efficiency{:} analyzing and optimizing large-scale {Google} {TPU} systems with {ML} productivity goodput.
\newblock 2025,  arXiv preprint arXiv: 2502.06982

\bibitem{Ascendllm_arxiv_2024}
Song M, Tang X, Hou F, Li~J, Wei W, Ma~Y, Xiao R, Si~H, Jiang D, Yin S, Hu~Y, Long G.
\newblock Tackling the dynamicity in a production {LLM} serving system with {SOTA} optimizations via hybrid prefill/decode/verify scheduling on efficient {Meta-kernels}.
\newblock 2024,  arXiv preprint arXiv: 2412.18106

\bibitem{Ascend_bench_1}
Wang B, Yang C, Zhu R, Liu X, Zhou M, Zheng N.
\newblock Analysis of performance and optimization in {MindSpore} on {Ascend} {NPUs}.
\newblock In: Proceedings of the 29th {IEEE} International Conference on Parallel and Distributed Systems.
\newblock 2023,  1701--1708

\bibitem{Ascend_bench_3}
Wu~R, Li~M, Li~H, Chen T, Tian X, Xu~X, Zhou B, Chen J, An~H.
\newblock Machine learning-enabled performance model for {DNN} applications and {AI} accelerator.
\newblock In: Proceedings of the 2022 {IEEE} 24th Int Conf on High Performance Computing {\&} Communications; 8th Int Conf on Data Science {\&} Systems; 20th Int Conf on Smart City; 8th Int Conf on Dependability in Sensor, Cloud {\&} Big Data Systems {\&} Application.
\newblock 2022,  25--34

\bibitem{ascend-cs24}
Xue W, Yang K, Liu Y, Fan D, Xu~P, Tian Y.
\newblock Unlocking high performance with low-bit {NPUs} and {CPUs} for highly optimized {HPL-MxP} on {Cloud Brain} {II}.
\newblock In: Proceedings of International Conference for High Performance Computing, Networking, Storage, and Analysis.
\newblock 2024, ~82

\bibitem{reddit}
Hamilton W~L, Ying Z, Leskovec J.
\newblock Inductive representation learning on large graphs.
\newblock In: Proceedings of the 31st International Conference on Neural Information Processing Systems.
\newblock 2017,  1024--1034

\bibitem{orkut_amazon}
Yang J, Leskovec J.
\newblock Defining and evaluating network communities based on ground-truth.
\newblock In: Proceedings of the 12th {IEEE} International Conference on Data Mining.
\newblock 2012,  745--754

\bibitem{10.1145/1772690.1772756}
Leskovec J, Huttenlocher D~P, Kleinberg J~M.
\newblock Predicting positive and negative links in online social networks.
\newblock In: Proceedings of the 19th International Conference on World Wide Web.
\newblock 2010,  641--650

\bibitem{products}
Ramezani M, Cong W, Mahdavi M, Sivasubramaniam A, Kandemir M~T.
\newblock {GCN} meets {GPU:} decoupling ``when to sample" from ``how to sample".
\newblock In: Proceedings of the 34th International Conference on Neural Information Processing Systems.
\newblock 2020

\bibitem{Livejournal}
Leskovec J, Lang K~J, Dasgupta A, Mahoney M~W.
\newblock Community structure in large networks{:} natural cluster sizes and the absence of large well-defined clusters.
\newblock Internet Mathematics, 2009, 6(1): 29--123

\bibitem{ascend-martrix}
Tang Y, Zhou H, Ji~Z, Wang C.
\newblock {Cube-fx:} mapping {Taylor} expansion onto matrix multiplier-accumulators of {Huawei} {Ascend} {AI} processors.
\newblock IEEE Transactions on Parallel and Distributed Systems, 2025, 36(6): 1115--1129

\bibitem{GPGPU-iscas24}
Wang C, Song P, Zhao H, Zhang F, Wang J, Zhang L.
\newblock High-utilization {GPGPU} design for accelerating {GEMM} workloads: {an} incremental approach.
\newblock In: Proceedings of 2024 {IEEE} International Symposium on Circuits and Systems.
\newblock 2024,  1--5

\bibitem{gnnhbmsample-hpec24}
Gui Y, Wu~Q, Yuan W, Liang H, Wang X, Jin X.
\newblock {HBM-based} hardware accelerator for {GNN} sampling and aggregation.
\newblock In: Proceedings of 2024 {IEEE} High Performance Extreme Computing Conference.
\newblock 2024,  1--7

\bibitem{hongtu-corr23}
Wang Q, Chen Y, Wong W, He~B.
\newblock {HongTu:} scalable full-graph {GNN} training on multiple {GPUs} (via communication-optimized {CPU} data offloading).
\newblock 2023,  arXiv preprint arXiv: 2311.14898

\bibitem{Quiver_corr_2023}
Tan Z, Yuan X, He~C, Sit M, Li~G, Liu X, Ai~B, Zeng K, Pietzuch P~R, Mai L.
\newblock {Quiver:} supporting {GPUs} for low-latency, high-throughput {GNN} serving with workload awareness.
\newblock 2023,  arXiv preprint arXiv: 2305.10863

\bibitem{pca}
Wang D, Xu~J.
\newblock Principal component analysis in the local differential privacy model.
\newblock In: Proceedings of the 28th International Joint Conference on Artificial Intelligence, {IJCAI} 2019.
\newblock 2019,  4795--4801

\bibitem{nextdoor_eurosys21}
Jangda A, Polisetty S, Guha A, Serafini M.
\newblock Accelerating graph sampling for graph machine learning using {GPUs}.
\newblock In: Proceedings of the 16th European Conference on Computer Systems.
\newblock 2021,  311--326

\bibitem{pagraph}
Lin Z, Li~C, Miao Y, Liu Y, Xu~Y.
\newblock {PaGraph:} scaling {GNN} training on large graphs via computation-aware caching.
\newblock In: Proceedings of the 11th ACM Symposium on Cloud Computing.
\newblock 2020,  401--415

\bibitem{ai2025neutronascend}
Ai~X, Zhang B, Wang Q, Zhang Y, Yuan H, Gong S, Yu~G.
\newblock Neutronascend: Optimizing gnn training with ascend ai processors.
\newblock ACM Transactions on Architecture and Code Optimization, 2025

\bibitem{ascend_Verrocchio_jpdc_23}
Tang Y, Wang C.
\newblock Performance modeling on {DaVinci} {AI} core.
\newblock Journal of Parallel and Distributed Computing, 2023, 175: 134--149

\end{thebibliography}

\begin{biography}{ckf.jpg}
Kefu Chen is currently a master's student in computer science at Northeastern University, China. His major research interests include acceleration of graph computing and learning system on emerging hardware.
\end{biography}

\vspace{0.2in}
    
\begin{biography}{ax}
Xin Ai is currently working toward the PhD degree in computer science with Northeastern University. His research interests include parallel and distributed graph computing and learning system.
\end{biography}

\vspace{0.2in}

\begin{biography}{wqg}
Qiange Wang received the PhD degree in computer science from Northeastern University, China, in 2022. He is currently a postdoctoral research fellow at the National University of Singapore. His research interests include distributed graph processing, learning, and management systems.
\end{biography}

\vspace{0.2in}

\begin{biography}{zyf}
Yanfeng Zhang received the PhD degree in computer science from Northeastern University, China, in 2012. He is currently a professor with Northeastern University, China. His research consists of distributed systems and big data processing. He has published many papers in the above areas. His paper in SoCC 2011 was honored with ``Paper of Distinction”.
\end{biography}

\vspace{0.2in}

\begin{biography}{yg}
Ge Yu received the PhD degree in computer science from the Kyushu University of Japan, in 1996. He is now a professor with Northeastern University, China. His current research interests include distributed and parallel systems, cloud computing, big data management, and blockchain techniques and systems. He has published more than 200 papers in refereed journals and conferences. 
He is a Fellow of CCF and a Senior Member of IEEE and ACM.
\end{biography}

\end{document}